\documentclass[review]{elsarticle}

\usepackage{graphicx}
\usepackage{amssymb}
\usepackage{amsthm}
\usepackage{amsmath}
\usepackage{bm}
\usepackage{color}
\usepackage{tabularx} 
\usepackage{booktabs} 
\usepackage{siunitx} 
\usepackage[caption=false]{subfig} 
\usepackage{capt-of}
\usepackage{natbib}
\usepackage{float}
\usepackage{pgfplots}
\usepackage{algorithm,algpseudocode}  
\usepgfplotslibrary{external}
\usepackage{pgfplots} 
\pgfplotsset{compat=1.16}

\usepackage{geometry}
\usepackage{siunitx} 
\usepackage{booktabs} 

\usepackage[linktocpage]{hyperref} 			

\newcommand{\reals}{\mathbb{R}}
\newcommand{\dataset}{\mathcal{D}}
\newcommand{\minibatch}{\mathcal{B}}
\newcommand{\normaldist}{\mathcal{N}}
\newcommand{\condbar}{\,|\,}  
\newcommand{\process}{\mathcal{P}}

\newcommand{\expectationp}[2][]{\mathbb{E} \ifx #1 \undefined \else _{#1} \fi \left[#2\right]}
\newcommand{\variancep}[2][]{\mathsf{Var} \ifx #1 \undefined \else _{#1} \fi \left[#2\right]}

\newcommand\ph{$\phantom{1}$}

\usepackage{lineno}
\modulolinenumbers[2]

\biboptions{authoryear}

\journal{Elsevier for possible publication}

\begin{document}

\begin{frontmatter}

\title{Passive learning to address nonstationarity in virtual flow metering applications}

\address[label1]{Department of Engineering Cybernetics, NTNU, O. S. Bragstads plass 2D, 7034 Trondheim, Norway}
\address[label2]{Solution Seeker AS, Rådhusgata 24, Oslo, Norway}
\cortext[cor1]{I am corresponding author}

\author[label1,label2]{Mathilde Hotvedt\corref{cor1}}
\ead{mathilde.hotvedt@ntnu.no / mathilde.hotvedt@gmail.com, +47 994 717 50}

\author[label1,label2]{Bjarne Grimstad}
\ead{bjarne.grimstad@solutionseeker.no}


\author[label1]{Lars Imsland}
\ead{lars.imsland@ntnu.no}

\begin{abstract}
Steady-state process models are common in virtual flow meter applications due to low computational complexity, and low model development and maintenance cost. Nevertheless, the prediction performance of steady-state models typically degrades with time due to the inherent nonstationarity of the underlying process being modeled. Few studies have investigated how learning methods can be applied to sustain the prediction accuracy of steady-state virtual flow meters. This paper explores passive learning, where the model is frequently calibrated to new data, as a way to address nonstationarity and improve long-term performance. An advantage with passive learning is that it is compatible with models used in the industry. Two passive learning methods, periodic batch learning and online learning, are applied with varying calibration frequency to train virtual flow meters. Six different model types, ranging from data-driven to first-principles, are trained on historical production data from 10 petroleum wells. The results are two-fold: first, in the presence of frequently arriving measurements, frequent model updating sustains an excellent prediction performance over time; second, in the presence of intermittent and infrequently arriving measurements, frequent updating in addition to the utilization of expert knowledge is essential to increase the performance accuracy. The investigation may be of interest to experts developing soft-sensors for nonstationary processes, such as virtual flow meters. 
\end{abstract}


\end{frontmatter}


\section{Introduction}\label{sec:introduction}
Many real-world, physical processes are nonstationary \citep{Sayed-Mouchaweh2012}. To various degrees, process conditions and properties change with time. Nevertheless, a common assumption in process modeling is time independence, leading to stationary, or steady-state, models \citep{Granero-Belinchon2019}. Several arguments militate for the utilization of steady-state models. Firstly, many processes are slowly time-varying making the stationary assumption reasonable for short-term applications. Secondly, steady-state models typically reduce the cost of model development and maintenance \citep{Solle2016}. Thirdly, these models are often less computationally heavy, which can increase the suitability in real-time control and optimization applications \citep{ModSim}. On the other hand, the performance of steady-state models in nonstationary conditions typically degrade with time and necessitates algorithms that improve the handling of nonstationarity.

\begin{figure}[ht]
	\centering
	\includegraphics[width=1.0\linewidth]{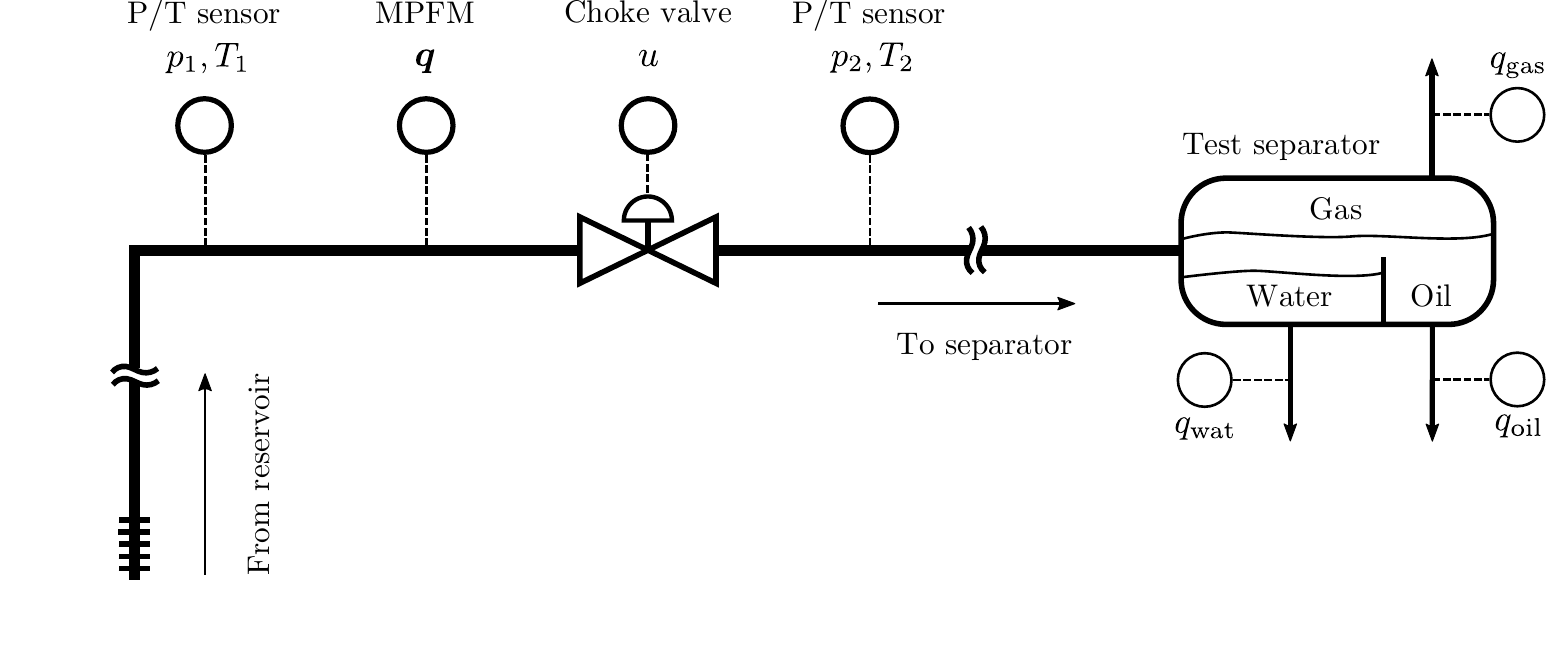}
	\caption{A simplified illustration of the petroleum production system with typical sensor placements. A multiphase flow meter (MPFM) measures the phasic flow rates through the choke valve. Measurements of the phasic flow rates can also be obtained when the well is tested, using, for instance, a test separator.}
	\label{fig:well_sensors}
\end{figure}

Virtual flow metering (VFM) is a soft-sensor technology that utilizes process models for continuous prediction of the multiphase flow rate at key locations in a petroleum asset \citep{Toskey2012}. In Figure \ref{fig:well_sensors}, a simplified illustration of the production system for one petroleum well is given along with typically available sensor measurements for well-equipped wells. 
A multiphase flow meter (MPFM) measures the phasic flow rates, $\bm{q} = [q_{\text{gas}}, q_{\text{oil}}, q_{\text{water}}]$, through the production choke valve. Under well-testing, the phasic flow rates can be measured using the test separator. The total multiphase flow rate through the production system is $Q = q_{\text{gas}} + q_{\text{oil}} + q_{\text{water}}$. A typical application of VFM is as a back-up to the MPFM in case of failure \citep{Varyan2015}. 

The underlying process of the VFM comprises the reservoir, wells, pipelines, and processing facility. This process is nonstationary with time-varying process conditions and properties \citep{Guo2007}. The multiphase flow rate through the production system has a dynamic nature with both fast and slow transients. Fast transients occur with control changes, which induce pressure waves through the system, such as the opening of the choke valve \citep{NodalAnalysis2015}. These are in the time range of minutes to hours. Slow transients are caused by the reservoir being depleted with time, which in turn results in a pressure declination in the production system and a decreased production flow rate \citep{Foss2018}. These occur in a time range of months to years, dependent on the size of the reservoir. Furthermore, as the petroleum asset ages, technologies such as artificial lift with gas or water are applied to improve production. Other sporadic changes such as maintenance tasks will also induce transient process behavior. Hence, the natural approach to VFM is nonstationary models. Several commercial VFMs such as Olga and LedaFlow are nonstationary \citep{Amin2015}, and other examples exist in literature \citep{Holmos2011, Jordanou2017}. On the other hand, due to the slow dynamics of the reservoir, steady-state reservoir conditions for a certain time interval can often be assumed \citep{Shippen2012}. Furthermore, considering the inherent complex multiphase flow characteristics, which make it challenging to develop and solve nonstationary VFMs, steady-state VFMs are the most common approach in literature \citep{Bikmukhametov2019}, both for physics-based models \citep{Shippen2012, Varyan2015} and machine learning (ML) models \citep{AlQutami2017a, AlQutami2017b, AlQutami2017c, AlQutami2018, Bikmukhametov2020a, Grimstad2021}. Nevertheless, studies show that steady-state VFM models should be updated or recalibrated in time to provide adequate long-term prediction accuracy \citep{Sandnes2021, Hotvedt2021}. 
Several model learning methods exist that attempt to account for nonstationarity without imposing temporal dependencies in the model. The learning methods can be divided into an active or passive method \citep{Ditzler2015}. In passive learning, the process is assumed to be continuously changing and the model is routinely updated with access to new measurements. In active learning, statistical tests are used to detect significant changes in the process conditions, whereupon model updating is initiated. 

For the VFM application, it is not uncommon that new observations arrive infrequently, for example, twice a year or at the most once per month under well-testing \citep{Monteiro2020}. In such an event, active learning is redundant as the process conditions and properties are likely to have changed significantly during the elapsed time, and the model should be updated with each new measurement. For assets with access to continuous flow rate measurements, such as MPFM measurements, the VFM models would likely benefit from updating using these measurements in between well-tests. Nevertheless, in industry, even with frequent access to new measurements, model learning can occur intermittently due to limited resources or manual, non-systematic workflows \citep{Koroteev2021}.

To the authors' knowledge, no studies have investigated the influence of the update frequency on sustaining the prediction accuracy of steady-state VFM models over time, hence, obtaining a high long-term performance. This research contributes in this direction by examining two passive learning methods: periodic batch learning and online learning. Six VFM models are developed for the petroleum production choke valve in 10 petroleum wells on Edvard Grieg, an asset on the Norwegian Continental Shelf \citep{EG}. Real production data spanning five years are used in the development. The long-term predictive performance is expected to increase with the frequency of which the models are updated. The best performance is expected from online learning, for which the models are updated with every new measurement. For periodic batch learning, the performance is expected to drop as the frequency is lowered. The rest of the article is structured the following way: section \ref{sec:nonstationary-env} presents relevant theory for steady-state modeling of processes in nonstationary conditions. Thereafter, Section \ref{sec:data-models} describes the available data and the VFM model types. In Section \ref{sec:numerical-study}, the numerical study examining the learning methods is described and results visualized and discussed. Lastly, Section \ref{sec:conclusions} gives concluding remarks. 

\section{Steady-state modeling in nonstationary conditions}\label{sec:nonstationary-env}
Consider a stream of observations $S = \{(\bm x_1, y_1), (\bm x_2, y_2), \ldots, (\bm x_t, y_t), \ldots \}$, where $\bm{x}_t \in \reals^d$ represents measured process conditions and $y_t \in \reals$ a (dependent) target variable at time $t$. In general, the set $S$ can be thought of as a realization of a stochastic process $\process$ governed by a generative model \citep{Oliveira2021}
\begin{equation}\label{eq:generative-model}
    p_t(\bm{x}, y) = p_t(y \mid \bm{x})p_t(\bm{x}).
\end{equation} 
In \eqref{eq:generative-model}, $p_t(\bm{x})$ is the marginal distribution of the process conditions, and $p_t(y \mid \bm{x})$ is the conditional distribution of the target, both at time  $t$. The index $t$ indicates that the distributions may be time-variant, and therefore $\process$ may be nonstationary. 


In real-time applications of machine learning, like data-driven virtual flow metering, it is natural to develop models on historical data and test the model performance on future data. Collect in $\dataset_{a:b} = \{(\bm{x}_t, y_t)\}_{t=a}^{b}$ the sequence of observations with $t \in [a, b]$, and in $\dataset_{a} = \{(\bm{x}_t, y_t)\}_{t=a}$ the single observation at $t = a$. For a model to be developed at time $t=T$, the training dataset is denoted by $\dataset^{tr} = \dataset_{1:T}$ and the test dataset by $\dataset^{tr} = \dataset_{T+1:\infty}$.

Many machine learning models and algorithms are based upon the assumption that the training and test dataset originate from the same probability distribution; the data points in $S$ are independent and identically distributed (i.i.d.) \citep{Hastie2009}. When the stochastic process $\process$ in \eqref{eq:generative-model} is nonstationary, the i.i.d. assumption is invalidated, as a dataset shift can occur when moving from the training phase to the test phase. In the following, different types of dataset shifts are explored, and suitable learning methods to alleviate the effect of nonstationarity on predictive performance are discussed.

\subsection{Dataset shifts}
When $\process$ is nonstationary, the joint probability distribution can shift in time resulting in $p_{t}(\bm{x}, y) \neq  p_{t + \tau}(\bm{x}, y)$ for an arbitrary lapse $\tau>0$ in time. Using the model in \eqref{eq:generative-model} two types of dataset shifts, also called concept drifts, can occur in time: virtual and real drift\footnote{Other naming conventions for virtual drift are virtual concept drift, covariate shift, or input drift. Real drift is also known as real concept drift or output drift.} \citep{Quinonero-Candela2009, Ditzler2015}. With virtual drift, the marginal distribution shifts in time. That is, $p_{t}(\bm{x}) \neq p_{t + \tau}(\bm{x})$ for $\tau > 0$.
With real drift, the conditional distribution shifts in time, such that $p_{t}(y \mid \bm{x}) \neq p_{t + \tau}(y \mid \bm{x})$ for $\tau > 0$. Real and virtual drift may happen separately or simultaneously, in any case shifting the joint distribution with time. Notice, in \citep{Quinonero-Candela2009}, several other specialized forms of dataset shifts are discussed. 

\begin{figure}[ht]
\centering
\subfloat[Virtual drift. The marginal distribution $p(x)$ changes from one time step to another causing a change $p(y \mid x)$. The parameters remain unchanged.]{
\includegraphics[width=0.45\textwidth]{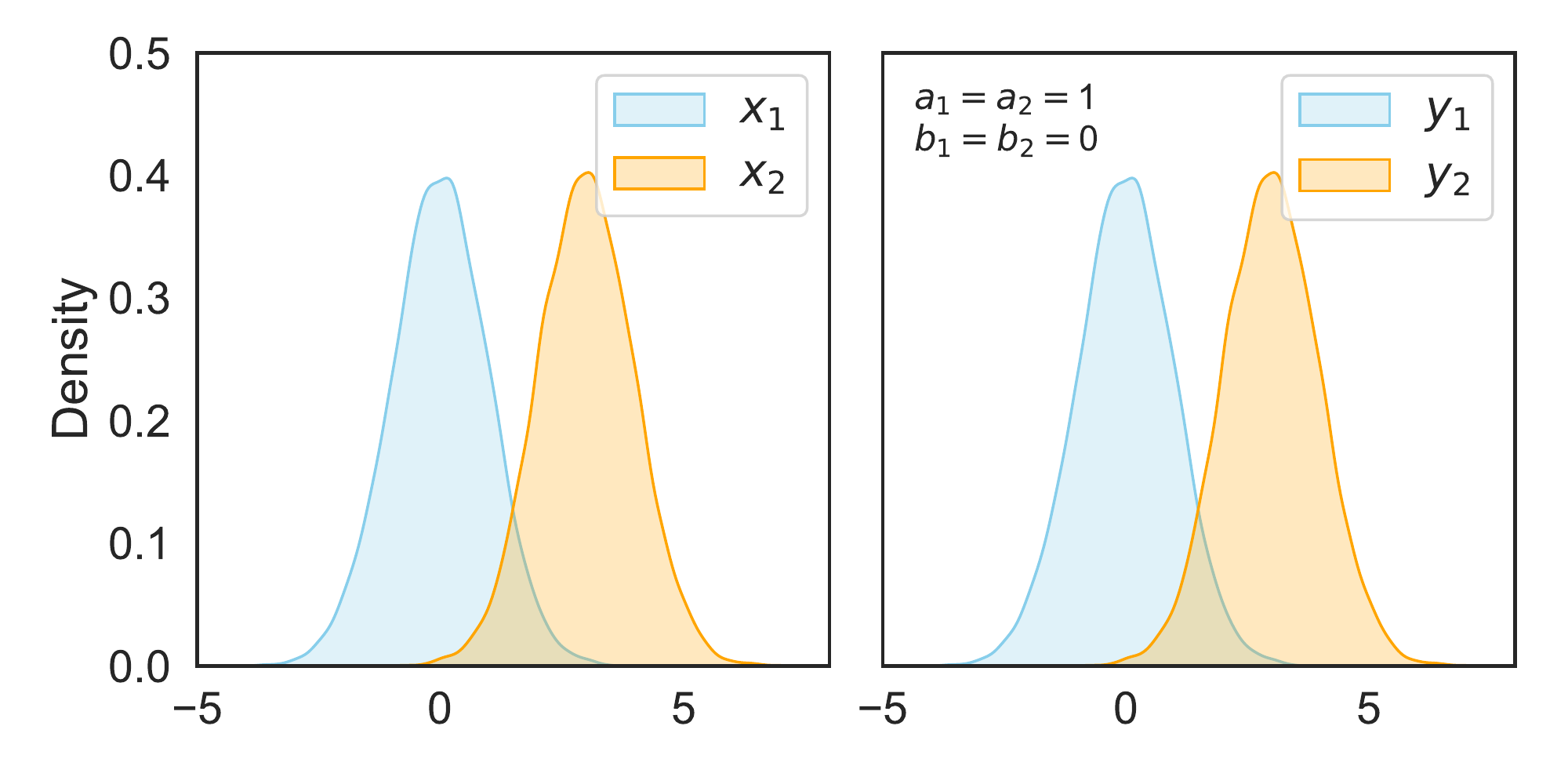}
\label{fig:virtual-drift}
}
\hfill
\subfloat[Real drift. The marginal distribution $p(x)$ does not change but $p(y \mid x)$ changes as a consequence of the changed model parameters.]{
\includegraphics[width=0.45\textwidth]{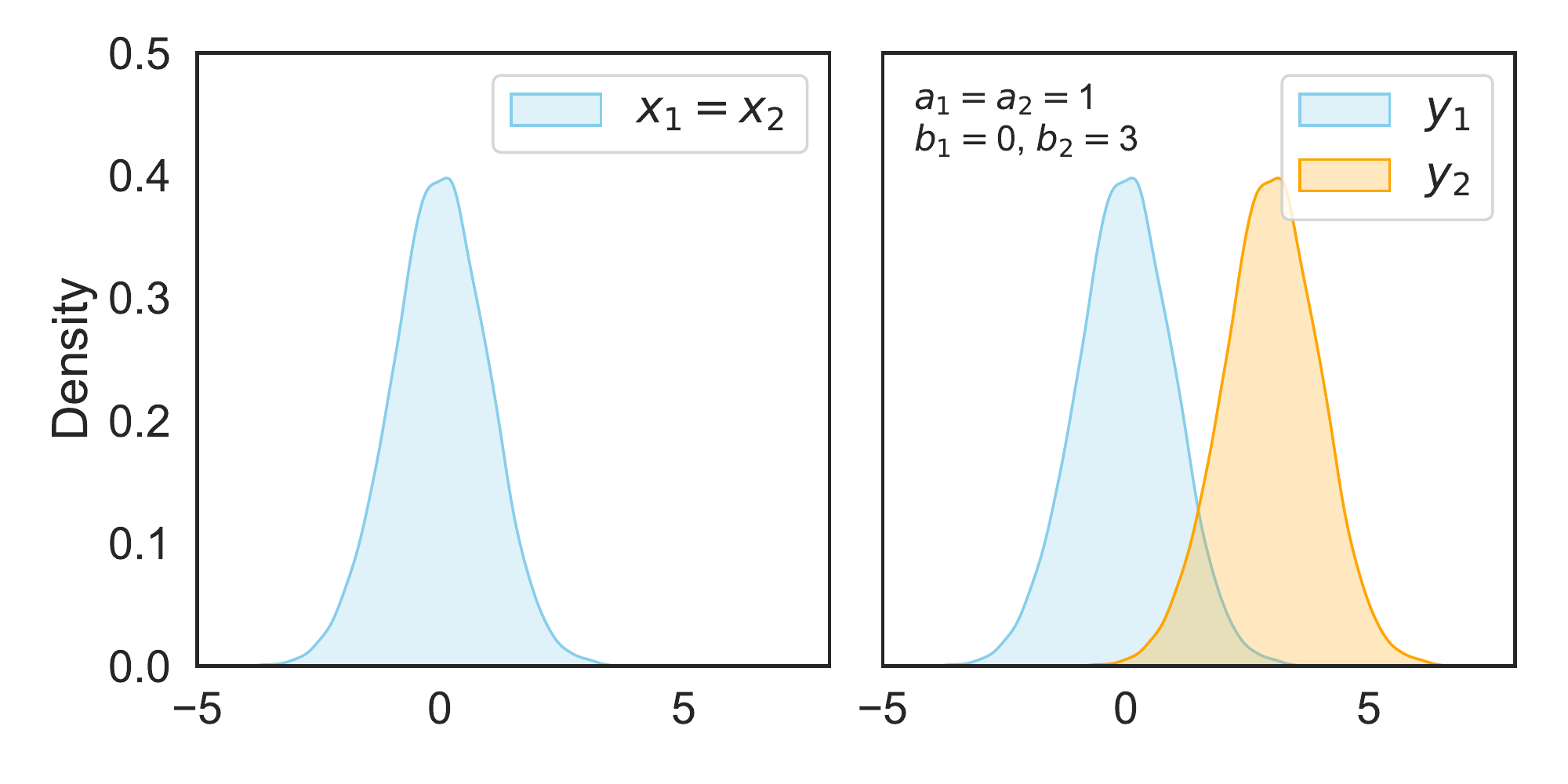}
\label{fig:real-drift}
}
\caption{Dataset shifts illustrated with a) virtual drift and b) real drift.}
\label{fig:nonstationarity-example}
\end{figure}

As an example, consider a process with a conditional distribution 
\begin{equation}\label{eq:linear-model}
    y_t = a_t x_t + b_t,
\end{equation}
with parameters $\bm{\theta}_t = \{a_t,b_t\}$. Two subsequent time instances $t=1$ and $t=2$ are examined. The input at $t=1$ is sampled from $p_{1}(x) \sim \normaldist(0, 1)$. At $t=2$, the mean changes such that $p_{2}(x) \sim \normaldist(3, 1)$.  If the model parameters remain unchanged, this is virtual drift, and the response in $y$ changes only as a consequence of changes in the marginal distribution. The scenario is illustrated in Figure \ref{fig:virtual-drift}. In another scenario, consider the input distribution to remain unchanged, but the $b$ parameter of the model to change from $b_{1}=0$ to $b_{2}=3$. The parameter change causes the conditional distribution in \eqref{eq:linear-model} to change, thereby causing real drift, illustrated in Figure \ref{fig:real-drift}. Notice, in the two scenarios, virtual and real drift cannot be distinguished by analyzing $y$ only. 

Virtual drift is commonly seen in the VFM application. For example, in time with the reservoir being depleted the pressure through the production system decreases. At the same time, in the early life of a petroleum asset, the production engineers can often increase the choke openings to maintain a constant production rate, also called plateau production \citep{NodalAnalysis2015}. The VFM application can also experience real drift. Substantial mechanical wear of the equipment in the well can occur with time, for instance, due to sand production, and can result in a change in the flow rate even for unchanged process conditions. It is believed that virtual drift is the major cause of observed dataset shifts in VFMs. However, as Figure \ref{fig:nonstationarity-example} illustrates, it can be difficult to separate between the two types of drifts. 

The next section discusses the impact that dataset shifts can have on steady-state VFM models.

\subsection{Parameter estimation of steady-state models}\label{sec:classical-approach}
A common approach to steady-state modeling is to use an inductive method to learn an approximation of the conditional distribution $p_t(y \condbar \bm x)$ in \eqref{eq:generative-model} from a fixed set of steady-state observations $\dataset_{1:T}$. A typical form of the approximation is 
\begin{equation}\label{eq:steady-state-model}
    \hat{y}_t = f_{\bm{\theta}}(\bm{x}_t) + \epsilon_t, \quad \epsilon_t \sim \normaldist (0, \sigma_{\epsilon}^2),
\end{equation}
where $f_{\bm{\theta}}$ is a parametric model of the mean, with parameters $\bm \theta$, and $\epsilon_t$ is a homoscedastic noise term. The model in \eqref{eq:steady-state-model} is a steady-state model since $\hat{y}_t$ is conditioned on $\bm x_t$, and the parameters $\bm{\theta}$ and $\sigma_\epsilon$ are time-invariant. The i.i.d. assumption is thus used. Note that, the resulting model is steady-state even though the data used to learn the model originate from a nonstationary process. 

Conditional models, like the steady-state model in \eqref{eq:steady-state-model}, are commonly trained using maximum a posteriori (MAP) estimation. In MAP estimation, the mode of the posterior distribution $p(\bm{\theta} \mid \dataset) \propto  p(\dataset \mid \bm{\theta})p(\bm{\theta})$ is maximized. Here, the likelihood $p(\dataset \mid \bm{\theta})$ is given by \eqref{eq:steady-state-model} and $p(\bm{\theta})$ is a prior on the $\bm \theta$ parameters. For a normal prior, $\theta_{i} \sim \normaldist(\mu_{i}, \sigma_{i}^2)$, $i = 1,..,N_{\theta}$, the optimization problem can be expressed as follows:

\begin{equation}\label{eq:objective-function}
\begin{aligned}
    \bm{\hat{\theta}} &= \arg \max_{\bm{\theta}} ~ \log p( \dataset \condbar \bm{\theta}) + \log p(\bm{\theta}) \\
    &= \arg \min_{\bm{\theta}} ~ \sum_{i=1}^{N} \frac{1}{\sigma_{\epsilon}^2} \left(y_i - \hat{y}_i\right)^2 + \sum_{i=1}^{N_{\theta}} \frac{1}{\sigma_{i}^2}\left(\theta_{i} - \mu_{i}\right)^2.
\end{aligned}
\end{equation}
where $N$ is the number of data points in the training dataset. From \eqref{eq:objective-function}, it is seen that MAP estimation is a trade-off between minimizing the squared errors and parameter deviation away from its respective mean value $\mu_{i}$. By multiplying the objective function by $\sigma_{\epsilon}^{2}/N$, the equivalence of MAP estimation to the familiar minimization of mean squared error with $\ell_2$-regularization is obtained \citep{Goodfellow2016}. 

In the machine learning domain, \eqref{eq:objective-function} is commonly optimized by first-order gradient descent methods \citep{Bishop2006}. These methods update the parameters iteratively according to the following scheme:
\begin{equation}\label{eq:iterative-opt}
    \bm{\hat{\theta}}^{(k+1)} = \bm{\hat{\theta}}^{(k)} - \gamma^{(k)} \mathcal{M}(\minibatch,  \bm{\hat{\theta}}^{(k)}), \; k = 1,..,E
\end{equation}
where $E$ is the number of iterations or steps taken towards the optimal value, $\gamma$ is the learning rate or step-size, and $\mathcal{M}$ is the set of equations calculating the step direction. The $\minibatch$ is a set of observations extracted from the training dataset and can be in the range of one to all observations. Any parameter that is not included in $\bm{\theta}$ is called a hyperparameter, for instance, $\gamma$, $E$, and $|\minibatch|$. 

The above approach to steady-state modeling is susceptible to dataset shifts since the estimate (optimum) in \eqref{eq:objective-function} likely will change with time, resulting in poor test performance. When applied to VFM, for which the data is generated by a nonstationary process, both virtual and real concept drift will negatively influence the long-term predictive performance. A VFM performance that diminishes with time, has been documented in several publications \citep{Grimstad2021, Hotvedt2021, Sandnes2021}. In the following section, passive learning methods are discussed. These methods can be used to account for dataset shifts in steady-state modeling. 

\subsection{Passive learning for steady-state models}\label{sec:passive-learning}
In passive learning, the process $\process$ is assumed to be continuously changing with time, and model updating is routinely initiated regardless of whether or not dataset shifts occur. Two methods of passive learning are examined: online learning (OL) and periodic batch learning (PBL). 

\begin{figure}[ht]
	\centering
	\includegraphics[width=1.0\linewidth]{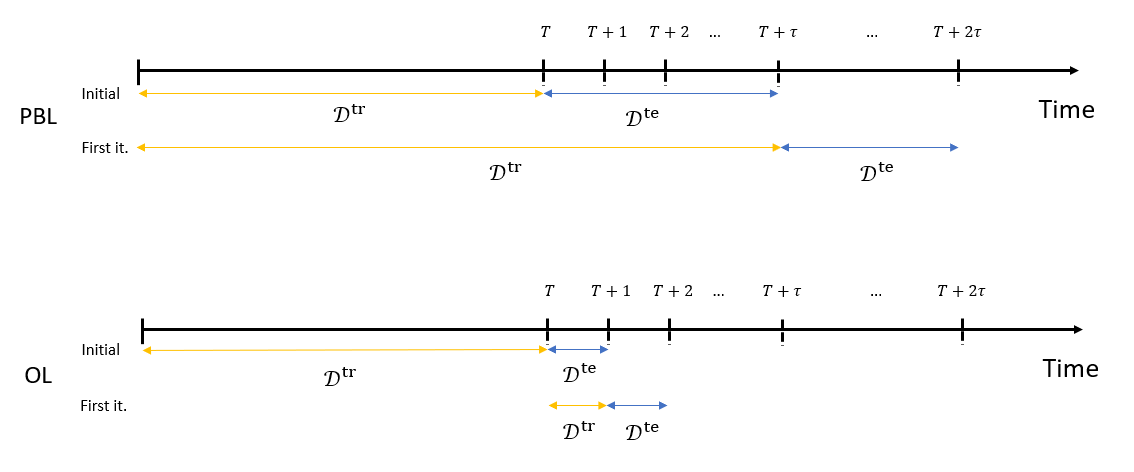}
	\caption{One iteration of the periodic batch learning and online learning update procedure after obtaining the initial parameter estimate. $\dataset^{\text{tr}}$ are training datasets used in the estimation problem and $\dataset^{\text{te}}$ are test datasets used to test the predictive capabilities of the model.}
	\label{fig:passive-learning-drawing}
\end{figure}

At time $t=T$, an initial parameter estimate is obtained from $\dataset^{\text{tr}} = \dataset_{1:T}$ using the approach in Section \ref{sec:classical-approach}. The estimated parameters are referred to as $\bm{\hat{\theta}}_T$, and the resulting steady-state model is given by
\begin{equation}\label{eq:steady-state-model-T}
    \hat{y}_t = f_{\bm{\hat{\theta}}_T}(\bm{x}_t) + \epsilon, \quad \epsilon \sim \normaldist (0, \sigma_{\epsilon}^2).
\end{equation}
From this point in time, the two learning methods can be applied. These are visualized in Figure \ref{fig:passive-learning-drawing} and are explained in the consecutive sections. 

\subsubsection{Periodic batch learning}\label{sec:periodic-batch-learning}
In periodic batch learning, the model in \eqref{eq:steady-state-model-T} is used to make predictions for $\tau>0$ time steps $\dataset^{\text{te}}_{T:\tau}$ before it is retrained at $t = T + \tau$. In retraining, the new parameters $\bm{\hat{\theta}}_{T+\tau}$ are estimated using all data observed at that time as training data $\dataset^{\text{tr}}_{1:T +\tau}$ and the approach in Section \ref{sec:classical-approach}. The procedure is repeated with a period of $\tau$, where the posterior parameter distribution can be described with 
\begin{equation}\label{eq:batch-learning}
    p(\bm{\theta} \condbar \dataset_{1:T+\tau}) \propto p( \dataset_{1:T+\tau} \condbar \bm{\theta})p(\bm{\theta}), 
\end{equation} 

An appropriate $\tau$ must be determined and can be accomplished by applying a change or shift detection algorithm offline on historical data. There exist much literature on shift detection algorithms, see for example \citet{Raza2015} and references therein. In this research, Hotelling's T-squared test for two multivariate, independent samples is used to investigate a null hypothesis stating that no virtual drift is present in the dataset. The algorithm for determining $\tau$ is described in \ref{app:update-frequency-pbl}.

\subsubsection{Online learning}\label{sec:online-learning}
In online learning, model updating occur for each new observation that arrives. However, the posterior distribution at the next time step is updated using only the current observation as the training data and the posterior distribution at the previous time step as the prior. For instance, at $t=T+1$:
\begin{equation}\label{eq:online-learning}
    p(\bm{\theta} \condbar \dataset_{1:T+1}) \propto p(\dataset_{T+1}\mid \bm{\theta}) p(\bm{\theta} \mid \dataset_{1:T}).
\end{equation} 
Mathematically, \eqref{eq:online-learning} can be derived as follows. With the approach in Section \ref{sec:classical-approach}, the posterior parameter distribution at $t=T+1$ is given by
\begin{equation}\label{eq:bayes-T+1}
    p(\bm{\theta} \condbar \dataset_{1:T+1}) = \frac{p( \dataset_{1:T+1} \condbar \bm{\theta})p(\bm{\theta})}{p(\dataset_{1:T+1})},
\end{equation} 
where $p(\dataset_{1:T+1})$ is the proportionality constant in Bayes' law. Applying the i.i.d. assumption, the likelihood function of the model and the evidence can be written as 
\begin{equation}\label{eq:idd-observations}
\begin{aligned}
    p(\dataset_{1:T+1} \mid \bm{\theta}) &= \prod_{t=1}^{T+1} p(\dataset_t \mid  \bm{\theta}) = p(\dataset_{1:T} \mid  \bm{\theta})p(\dataset_{T+1} \mid \bm{\theta}) \\ 
    p(\dataset_{1:T+1}) &= \prod_{t=1}^{T+1} p(\dataset_t) = p(\dataset_{1:T} )p(\dataset_{T+1}),
\end{aligned}
\end{equation}
respectively. Note that, while the i.i.d. assumption is likely false for a nonstationary process, it is already used in steady-state modeling. Inserting \eqref{eq:idd-observations} in \eqref{eq:bayes-T+1}, the posterior parameter distribution at $t=T+1$ can be written as
\begin{equation}
    p(\bm{\theta} \condbar \dataset_{1:T+1}) = \frac{p(\dataset_{T+1} \mid \bm{\theta})}{p(\dataset_{T+1})} \cdot \frac{ p(\dataset_{1:T} \mid  \bm{\theta})p(\bm{\theta})}{p(\dataset_{1:T})} = \frac{p(\dataset_{T+1} \mid \bm{\theta})}{p(\dataset_{T+1})} \cdot p(\bm{\theta} \mid \dataset_{1:T})
\end{equation}
and \eqref{eq:online-learning} is obtained.  

An issue becomes apparent when deriving the MAP estimate for \eqref{eq:online-learning} 
\begin{equation}\label{eq:MAP-OL}
\begin{aligned}
    \bm{\hat{\theta}}_{T+1} &= \arg \max_{\bm{\theta}} \Big[\log p(\dataset_{T+1} \mid \bm{\theta}) + \log p (\bm{\theta} \mid \dataset_{1:T})\Big]. \\
\end{aligned}
\end{equation}
Ideally, the parameter estimation in the previous time step should have provided both the mean and the variance of the updated posterior parameter distribution $p(\bm{\theta} \mid \dataset_{1:T}) \sim \normaldist (\bm{\mu}_{T}, \bm{\Sigma}_{T})$. However, MAP estimation gives point estimates of the mode only. When the likelihood and prior is normal, an estimate of the mean $\bm{\mu}_{T} = \bm{\hat{\theta}}_{T}$ is obtained since the mode and mean coincides, but $\bm{\Sigma}_{T}$ remains unknown. Therefore, the second term in \eqref{eq:MAP-OL} cannot be calculated if MAP estimation is used in each time step. As discussed in Section \ref{sec:classical-approach}, this term is $\ell_2$-regularization of the parameters. According to \citep{Goodfellow2016}, for some cases, the algorithm early stopping has a similar effect as $\ell_2$-regularization. For linear models, the solution obtained with early stopping equals a solution with $\ell_2$-regularization where the regularization term is determined by the number of iterations and step-size in early stopping \citep{Santos1994}. Therefore, for the OL algorithm implemented in this research, the iterative optimization algorithm in \eqref{eq:iterative-opt} uses the posterior parameter estimate from the previous time step as a starting point but iterates only a few steps $k$ towards the optimal value. In such a sense, the approach is similar to an early stopping approach, and will to some degree include parameter regularization. 

\subsection{Comparison of periodic batch and online learning}
There are advantages and disadvantages to both passive learning methods. With OL, the model can quickly adapt to changes in process conditions. Further, as only new observations are used, old data may be discarded yielding low memory requirements. However, it has been shown that some machine learning models such as neural networks are prone to catastrophic forgetting when trained using OL \citep{Goodfellow2013, Kemker2018, Parisi2019}. Catastrophic forgetting is a situation where the model excessively overfits its parameters to new observations resulting in a decreased performance on previously seen observations. This situation occurs due to the stability-plasticity dilemma \citep{Wickliffe2005}. The neural network requires adequate plasticity to adapt to new patterns, but too much can cause the network to forget previously learned patterns. The reverse is true for stability. The stability-plasticity of the models is connected to the hyperparameters of the learning algorithms. With time, the optimal hyperparameters can change. This is a problem for OL in real-time applications as a hyperparameter search in each iteration can be infeasible, dependent on the frequency of arrival of new observations. Another potential issue for the OL is the required complex system integration. The method will require fast processing capabilities of new observations to account for erroneous sensor measurements, and model performance monitoring applications are a necessity to analyze model drift and catastrophic forgetting \citep{Ditzler2015}. Furthermore, the learning method must be automated as manual, although systematic, handling of model updating can be impractical in real-time due to limited resources. 

PBL addresses catastrophic forgetting as all available observations are used in model updating. Yet, using this method for each new observation can be impractical in real-time applications due to a larger training time caused by larger datasets \citep{Kemker2018}. Therefore, a longer period ($\tau$ in \eqref{eq:batch-learning}) between model retraining can be required and sudden shifts in the data can be missed. On the other hand, if the underlying process is slowly changing, a lower update frequency can be sufficient to capture dominant changes in process conditions. Correspondingly, a manual yet systematic handling of the learning method including measurement preprocessing, conducting a hyperparameter search, and the actual model learning can be more achievable in each iteration. For VFM applications, studies have indicated that the inclusion of too old data may be redundant and not improve the model performance significantly \citep{AlQutami2018, Grimstad2021}. Thus, a windowing strategy can be applied to discard redundant data \citep{Ditzler2015}.

\section{Data and models}\label{sec:data-models}
In this research, six different VFM model types are considered. The data used to develop the VFMs and examine the effect of the learning methods on the long-term prediction performance are real production data from 10 wells, W1-W10, on the Edvard Grieg asset \citep{EG}. The available data and the VFM model types are described in the below sections.

\subsection{Available data}\label{sec:data}
The available process data consists of observations from the $M=10$ wells indexed by $j \in \{1, \ldots, M\}$. The dataset of well $j$ is $\{ (\bm{x}_{t,j}, y_{t,j}) \}_{t=1}^{N_j}$, where $N_j$ is the number of observations, explanatory variables are $\bm{x}_{t, j} = (u, p_{1}, p_{2}, T_{1}, \eta_{\text{oil}}, \eta_{\text{gas}})_{t,j} \in \reals^6$, and target variables are $y_{t, j} = Q_{t,j} \in \reals$. The $\eta_{\text{oil}}$ and $\eta_{\text{gas}}$ are the fractions of oil and gas in the fluid mixture. Ideally, the fractions should be estimated using a different model, for instance, a wellbore model as in \citet{Kittilsen2014}. For simplification, the fractions are approximated using the measured phasic volumetric flows. Measurements of the target variable, the mixture volumetric flow rate, are from both well-tests conducted with a test separator, and from the multiphase flow meter in each well. Commonly, well-test measurements have higher accuracy than MPFM measurements as MPFM are prone to failure and drift over time \citep{Falcone2013}. The data from all wells is denoted by $\dataset$. 
\begin{figure}[ht!]
	\centering
	\includegraphics[width=1.0\linewidth]{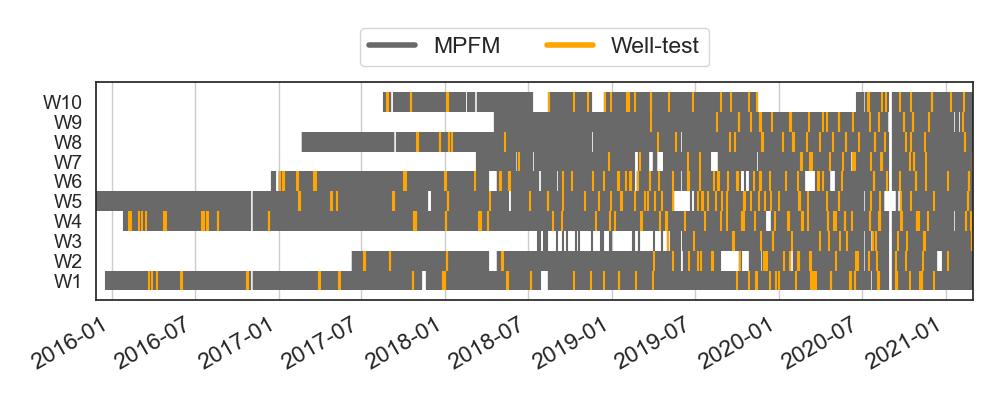}
	\caption{Visualization of the occurrence of observations for each well against time. Some wells have older historical observations than others. Both multiphase flow meter and well-test measurements are available. }
	\label{fig:dataset_visualization}
\end{figure}
Each of the datasets is generated using the processing technology in \citet{Grimstad2016}. This technology compresses the data by removing fast transients. However, slow transients can still be present. Further, the datasets are passed through a set of filters that remove undesired, illogical measurements, for instance, negative pressures or negative flow rate measurements. The wells have an unequal number of observations spanning a different time range, see Figure \ref{fig:dataset_visualization}. Some wells have historical observations back to 2016 while others have their first observations in late 2018. Further, there are periods where observations are lacking for some of the wells represented by white holes in the data in Figure \ref{fig:dataset_visualization}. Here, the well in question can have been shut down, or the sensors failed. In total, there are $26743$ observations from the 10 wells, spanning more than five years of production history. On average, there is less than one day between each measurement. The time between well-tests for a well is varying, with more than one year at the longest, and less than a day at the shortest.  

\subsection{Virtual flow meter models}\label{sec:models}
The six different VFM models considered range from machine learning, or data-driven, to physics-based, or mechanistic, models:
\begin{enumerate}
    \item A linear regression model (LR)
    \item A fully connected feed-forward neural network (NN)
    \item A multi-task learning model (MTL)
    \item A hybrid, gray-box error model (HEM)
    \item A hybrid, gray-box area function model (HAM)
    \item A mechanistic model (MM)
\end{enumerate}

There are advantages and disadvantages with all model types \citep{Solle2016, Hotvedt2021}. Mechanistic models are built from physical laws and require little process data in development. Yet, simplifications and assumptions are often necessary to make mechanistic models computationally feasible in real-time applications. Hence, model bias or process-model mismatch is typically encountered. Machine learning models are built from available data only and require no prior knowledge about the physics of the process. The capacity of machine learning models vary, where the NN is a typical model with high capacity and the LR a model with low capacity. High capacity models enable adaptation to arbitrarily complex physical relationships as long as these are reflected in the data, commonly reducing model bias. However, due to the inherent bias-variance trade-off of high capacity models, minimizing the bias results in higher variance \citep{Hastie2009}. Therefore, high capacity models are often influenced by poor quality data or data located in the small data regime, a situation not uncommon for the VFM application \citep{Grimstad2021}. Furthermore, higher variance typically decreases the generalization abilities to previously unobserved data, and such models can struggle if used in nonstationary environments where the process experiences dataset shifts. The hybrid models attempt to utilize knowledge from both the mechanistic and the data-driven modeling domain to preserve the advantages but diminish the disadvantages of both methods. 

The MTL models are somewhat different from the other model types. This model type enables learning from a plurality of wells, where each well presents a learning task. Instead of separately training a model for each well, which can be considered as single-task learning, the models are simultaneously trained. The advantage of using multi-task learning is two-fold. First, it allows for parameter sharing among models which can drastically improve data efficiency and predictive performance in the small-data regime. This is analogous to an MM whose equations are shared among wells. Second, compared to single-task learning, simultaneous training can lessen the effort and computational cost of developing models when the number of wells becomes large.

In the following sections, a mathematical description of the six VFM models is introduced. In addition to these, a benchmark model used to compare the performance of the models is described. 

\subsubsection{Benchmark model}\label{sec:models-benchmark}
A simple benchmark model predicts the flow rate to be the same as the last observed flow rate. Consider chronologically ordered observations $\{y_{1, j}, \ldots, y_{N_j, j}\}$ for well $j$ so that $y_{t, j}$ is observed after $y_{t-1, j}$. The prediction from the benchmark model is
\begin{equation}
    \hat{y}_{t, j} = y_{t-1, j}, \quad t=1,..,N_j, \quad j=1,...,M.
\end{equation}
Note that with this model the prediction is independent of the explanatory variables $\bm{x}_{t, j}$. Further, if the petroleum production is on plateau, resulting in each new observation deviating little from the previous, the benchmark model has the potential of high accuracy. 

\subsubsection{Linear regression model}\label{sec:models-LR}
The linear regression model fits a multidimensional line to the observed data. The functional form is given by $f_{\bm{\theta}}^{\text{(LR)}}: \reals^d \to \reals$ and is evaluated for a given $\bm{x}$ as 
\begin{equation}
    \hat{y} = \bm{w}^T\bm{x} + b.
\end{equation}
The model parameters consist of a weight vector $\bm{w}\in \reals^{d}$ and a bias $b \in \reals$, $\bm{\theta} = \{(\bm{w}, b)\}$. 

\subsubsection{Feed-forward neural network model}\label{sec:models-NN}
In general, the feed-forward neural network is a set of nonlinear regression lines. It has a functional form $f_{\bm{\theta}}^{\text{(NN)}}: \reals^d \to \reals$. For a neural network with $L$ hidden layers and one output layer, the parameters are $\bm{\theta} = \{(W^{(l)}, \bm{b}^{(l)})\}_{l=1}^{L+1}$, where $W^{(l)}$ and $\bm{b}^{(l)}$ are the weights and biases of layer $l$, respectively. The dimensions of $W^{(l)}$ and $\bm{b}^{(l)}$ determine the width of layer $l$.

In this work, the rectified linear unit (ReLU) activation function is used as the nonlinearity in the hidden layers \citep{Glorot2011}. The ReLU function is denoted by $a: \reals^d \to \reals^d$, $a(\bm{z})_i := \max(0, \bm{z}_i)$, where the $\max$ operator is applied element-wise for $i=1,\ldots,d$. This makes the neural network a set of piecewise linear regression lines. The evaluation of model $f_{\bm{\theta}}^{\text{(NN)}}(\bm{x})$ for a given $\bm{x}$ is
\begin{equation}
    \begin{aligned}
    \bm{z}^{(1)} &= \bm{x} \\
    \bm{z}^{(l+1)} &= a(W^{(l)} \bm{z}^{(l)} + \bm{b}^{(l)}), & l=1,\ldots,L \\
    \hat{y} &= W^{(L+1)} \bm{z}^{(L+1)} + \bm{b}^{(L+1)}. 
    \end{aligned}
    \label{eq:nn-model}
\end{equation}

\subsubsection{Multi-task learning model}\label{sec:models-MTL}
A MTL formulation introduces a new semantics of the model parameters compared to the NN in Section \ref{sec:models-NN}. Let $\bm{\alpha}$ denote parameters that are shared among tasks (here wells), and let $\bm{\beta}_j \in \reals^P$ be $P$ task-specific parameters for wells $j=1,\ldots, M$. The parameters of the MTL model for $M$ wells are collected in $\bm{\theta} = \{\bm{\alpha}, \bm{\beta}_1, \ldots, \bm{\beta}_M\}$.

When processing a data point $\bm{x}_{t,j}$ of well $j$, the model must select the corresponding task-specific parameters, $\bm{\beta}_j$. The selection can be made by introducing an encoding of tasks. Let $\bm{e}_j$ be an indicator vector of dimension $M$, with all zeros, except for a one in position $j$. By stacking the task-specific parameters in a matrix $B$ with columns $B_{*,j} = \bm{\beta}_j$, a selection can be made by performing the multiplication $\bm{\beta}_j = B \bm{e}_j$.

A simple MTL model is obtained by utilizing the selection mechanism described above. First, $\bm{\beta}_j$ is selected using the encoding $\bm{e}_j$. Next, $\bm{x}_{t,j}$ and $\bm{\beta}_j$ are fed through a residual neural network with shared parameters $\bm{\alpha}$. In this work, a residual neural network with pre-activation is used to allow for an identity mapping of the task-specific parameters \citep{He2016}. The resulting model is a simplified version of the MTL choke model introduced in \citep{Sandnes2021}.

The functional form of the MTL model is $f_{\bm{\theta}}^{\text{(MTL)}}: \reals^d \times \{0, 1\}^M \to \reals$, where the second argument is the task encoding vector. The evaluation of $f_{\bm{\theta}}^{\text{(MTL)}}(\bm{x}, \bm{e}_j)$ for a data point $\bm{x}$ of well $j$, is performed as follows:
\begin{equation}
    \begin{aligned}
    \bm{\beta}_j &= B \bm{e}_j, \\
    \hat{y} &= g_{\alpha}(\bm{x}, \bm{\beta}_j), \\
    \end{aligned}
    \label{eq:mtl-model}
\end{equation}
where $g_{\alpha}$ is a residual neural network with $L$ residual blocks given by
\begin{equation}
    \begin{aligned}
    \bm{z}^{(1)} &= W^{(0, 1)} \bm{x} + W^{(0, 2)} \bm{\beta}_j + \bm{b}^{(0)}, \\
    \bm{r}^{(l)} &= W^{(l, 2)} a(W^{(l, 1)} a(\bm{z}^{(l)}) + \bm{b}^{(l, 1)}) + \bm{b}^{(l, 2)}, & l=1,\ldots,L, \\
    \bm{z}^{(l+1)} &= \bm{r}^{(l)} + \bm{z}^{(l)}, & l=1,\ldots,L, \\
    \hat{y} &= W^{(L+1)} \bm{z}^{(L+1)} + \bm{b}^{(L+1)}.
    \end{aligned}
    \label{eq:mtl-model-resnet}
\end{equation}
The weights and biases in \eqref{eq:mtl-model-resnet} are collected in $\bm{\alpha}$ and are shared among the $M$ wells. These parameters can be learned from all the data in $\dataset$.

\subsubsection{Mechanistic model}\label{sec:models-MM}
The mechanistic choke model is taken from \citet{Sachdeva1986}. The equations are developed from the steady-state mass and momentum balance equations for one-dimensional flow along a streamline.  In short notation, the mechanistic model is given by $f_{\bm{\theta}}^{(\text{MM})}: \reals^d \to \reals$ with parameters $\bm{\theta} = \{\rho_{\text{oil}}, \rho_{\text{wat}}, \kappa, M_{\text{gas}}, p_{cr}, C_D \}$, and the equation for the volumetric flow rate through the choke is given by:
\begin{equation}\label{eq:MM}
\begin{aligned}
    \hat{y} &= Q = \frac{\dot{m}}{\rho_{SC}} \\
    &= \frac{C_DA_2(u)}{\rho_{SC}} \times \sqrt{2\rho_2^2p_1\left(\frac{\kappa}{\kappa -1}\eta_{\text{gas}}\left(\frac{1}{\rho_{\text{gas},1}}-\frac{p_r}{\rho_{\text{gas},2}}\right) + \left(\frac{\eta_{\text{oil}}}{\rho_{\text{oil}}} + \frac{\eta_{\text{wat}}}{\rho_{\text{wat}}}\right)(1 - p_r)\right)},
\end{aligned}
\end{equation}
Details regarding the model are found in \citet{Hotvedt2021}.

\subsubsection{Hybrid error model}\label{sec:models-error}
This model uses the mechanistic model in Section \ref{sec:models-MM} as a baseline but inserts a neural network as introduced in Section \ref{sec:models-NN} to capture the error between the mechanistic model output and measurements, or the process-model mismatch. The functional form of the model is given by $f_{\bm{\theta}}^{(\text{HEM})}: \reals^d \to \reals$ with parameters $\bm{\theta} = \{\bm{\theta}_{\text{MM}}, \bm{\theta}_{\text{NN}}\}$, where the physical model parameters are the same as given in Section \ref{sec:models-MM}: $\bm{\theta}_{\text{MM}}=\{\rho_{\text{oil}}, \rho_{\text{wat}}, \kappa, M_{\text{gas}}, p_{cr}, C_D \}$, and the neural network parameters are the weights and biases on each layer of the network as described in Section \ref{sec:models-NN}: $\bm{\theta}_{\text{NN}} = \{(W^{(l)}, \bm{b}^{(l)})\}_{l=1}^{L+1}$. The evaluation of HEM for a data point $\bm{x}$ is described by 
\begin{equation}
    \hat{y} = f_{\bm{\theta}}^{(\text{HEM})}(\bm{x}) = f_{\bm{\theta}_{\text{MM}}}^{(\text{MM})}(\bm{x}) + f_{\bm{\theta}_{\text{NN}}}^{(\text{NN})}(\bm{x})
\end{equation}

\subsubsection{Hybrid area function model}\label{sec:models-area}
This model also uses the mechanistic model in Section \ref{sec:models-MM} as a baseline. However, the mechanistic relation for the area function $A_2(u)^{(\text{MM})}$ is manipulated by multiplying with a neural network. This may be interpreted as replacing the discharge coefficient $C_D$ from the MM with a neural network. Accordingly, $f_{\bm{\theta}}^{(\text{HAM})}: \reals^d \to \reals$ with parameters $\bm{\theta} = \{\bm{\theta}_{\text{MM}}, \bm{\theta}_{\text{NN}}\}$, where  $\bm{\theta}_{\text{MM}}=\{\rho_{\text{oil}}, \rho_{\text{wat}}, \kappa, M_{\text{gas}}, p_{cr} \}$ and $\bm{\theta}_{\text{NN}} = \{(W^{(l)}, \bm{b}^{(l)})\}_{l=1}^{L+1}$. The evaluation of $f_{\bm{\theta}}^{(\text{HAM})}$ for data point $\bm{x}$ is as follows: 
\begin{equation}
\begin{aligned}
    A_2 &= A_2(u)^{(\text{MM})} \times f_{\bm{\theta}_{\text{NN}}}^{(\text{NN})}(\bm{x}) \\
    \hat{y} &= f_{\bm{\theta}}^{(\text{HAM})}(\bm{x}) = f_{\bm{\theta}_{\text{MM}}}^{(\text{MM})}(\bm{x}, A_2)
\end{aligned}
\end{equation}
Note, the complete vector of explanatory variables is used as input to the area function network and not just the choke opening $u$. This is due to the expectation of the effective flow area being dependent on the characteristics of the fluid flowing through the choke, which cannot be captured with just $u$.

\subsection{Prior parameter distribution}
All the VFM models except the benchmark model need specification of the prior parameter distributions $\theta_i \sim \normaldist(\mu_i, \sigma_i^2)$. For the data-driven model parameters $\bm{\theta}_{\text{NN}}$, He-initialization is utilized, which is recommended for neural networks with ReLU as activation function \citep{He2015}. For the mechanistic model parameters $\bm{\theta}_{\text{MM}}$, typical values for the mean $\mu_i$ is commonly known. For instance, a typical value for the density of freshwater is $1000 kg/m^3$. The variance may be estimated using the known bounds of the parameter in question. Details on prior parameter specification in gray-box models may be found in \citet{Hotvedt2021}. 

\section{Numerical study}\label{sec:numerical-study}
Online learning and periodic batch learning as described in Section \ref{sec:passive-learning}, are used to train the six models in Section \ref{sec:models}, for the 10 wells, using the data described in Section \ref{sec:data}. Pay in mind, all VFM models except the LR are implemented using the Python framework \textit{PyTorch} \citep{pytorch}. The LR is implemented with the Python framework \textit{scikit-learn} \citep{scikit-learn} using the stochastic gradient descent linear regressor to allow for training the model with online learning. As mentioned in Section \ref{sec:online-learning}, the $\ell_2$-regularization term cannot be calculated for the online learning method. However, for the hybrid and mechanistic models, an important factor is that the model parameters with a physical interpretation $\bm{\theta}_{\text{MM}}$ stay within feasible bounds. Therefore, $\ell_2$-regularization with the initial priors is applied for these parameters. 

The numerical study considers two cases. In Case 1, all available data, both MPFM and well-test measurements are utilized in training. The initial parameter estimate is obtained with historical data before the 1st of January 2019, while the data after this point in time is used to test the learning methods, see Figure \ref{fig:dataset_visualization}. This split of data is referred to as the initial split. In Case 2, the models are trained using well-test measurements only. To ensure a sufficient amount of training data, the initial split is applied on the 1st of January 2020, see Figure \ref{fig:dataset_visualization}. 

Two analyses are conducted before the learning methods can be applied: 1) estimation of the PBL update frequency and 2) a search for optimal hyperparameters in the learning methods. These analyses are given in Section \ref{sec:update-freq-estimation} and \ref{sec:hyperparameter-search}, respectively, and are applied on the initial training data. From the outcome of the analyses, the models are trained with the learning methods, and the result for the two cases is given in \ref{sec:results-case-1} and \ref{sec:results-case-2}, respectively.

\subsection{Update frequency estimation for periodic batch learning}\label{sec:update-freq-estimation}
\begin{figure}[!ht]
\centering
\includegraphics[width=1.0\linewidth]{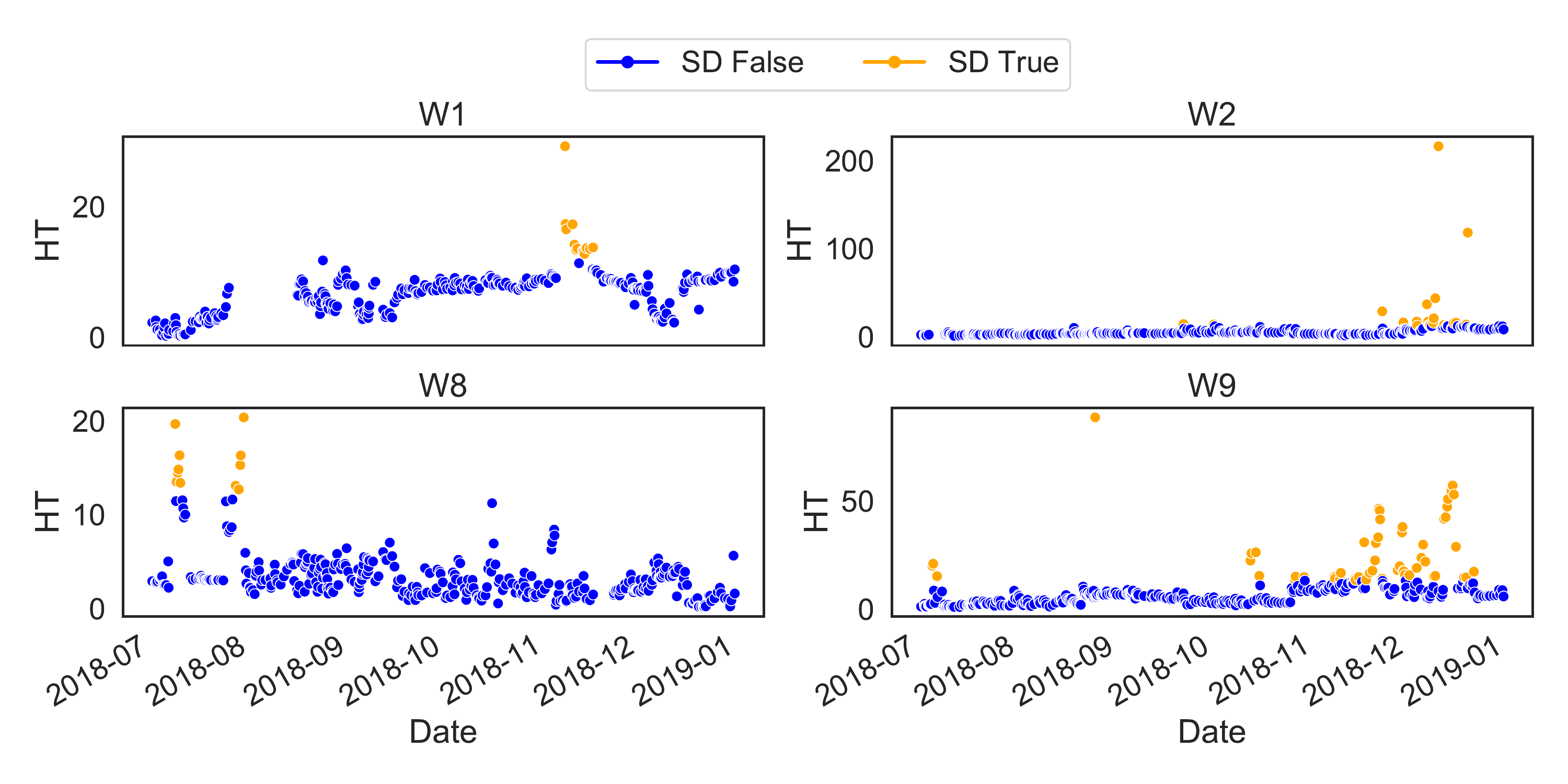}
\caption{The Hotelling's T-squared statistic with $\alpha = 0.05$ for each of the observation in the test dataset. The coloring indicates if the test observation is detected as a shift (SD: shift detection).}
\label{fig:HT-training}
\end{figure}

To estimate a suitable update frequency, Algorithm \ref{alg:update-freq} in \ref{app:update-frequency-pbl} with significance level $\alpha=0.05$ is used on the initial training data from Case 1. This data is split into two new datasets at time 01.07.2018. The six months of observations leading up to 01.01.2019 are used as the test dataset. From Figure \ref{fig:dataset_visualization}, it is seen that W3 does not have observations in the time range suggested. Therefore, the well is excluded from the analysis. In Figure \ref{fig:HT-training}, the $HT^2$ statistic for each observation in the test dataset is illustrated for four of the wells. The coloring indicates whether or not a shift is detected for the observation. 
W1 and W2 are the two wells of the nine examined with the longest period before a shift is detected, approximately after five months. W8 and W9 are the wells with the shortest period before a shift is detected, approximately after two weeks. Accordingly, different wells can have different optimal update frequencies, and it is likely to change during the lifetime of the petroleum asset. To simplify model learning, all wells are trained using the same update frequency. Therefore, the PBL is tested with a two weeks update frequency. The results are compared to a PBL with an update frequency of 6 months to examine the potential benefit of more frequent updating.  

\subsection{Hyperparameter search}\label{sec:hyperparameter-search}
For the periodic batch learning approach, a hyperparameter grid search for the learning rate is conducted testing $\gamma \in \{10^{-1}, 10^{-2}, 10^{-3}, 10^{-4}, 10^{-5}\}$. Early stopping is applied to determine the appropriate number of iterations $E$. For all VFM models except the LR, the optimizer Adam is applied. This optimizer have shown results in previous research on VFM modeling \citep{Hotvedt2020a, Hotvedt2020b, Hotvedt2021, Grimstad2021}. For the LR, Adam is not an option and the model is trained with SGD, yet, with the learning rate scheduler
\begin{equation}\label{eq:lr-scheduler}
    \gamma^{(k)} = \frac{\gamma^{(0)}}{k^{a}}
\end{equation}
where $\gamma^{(0)}$ is the initial learning rate, $k$ is the iteration number, and $a$ is a constant, see \eqref{eq:iterative-opt}. 

For the online learning approach, the hyperparameter grid search is extended to \newline $\gamma \in \{5\times 10^{-1}, 10^{-1}, 10^{-2}, 10^{-3}, 10^{-4}, 10^{-5}, 10^{-6}, 10^{-7}, 10^{-10}\}$. Pay in mind, $\gamma = 10^{-10}$ means close to negligible updating. As online learning processes only one sample at a time, early stopping cannot be applied. Therefore, the hyperparameter search includes experimentation with the number of iterations $E\in\{1, 10, 20\}$. For all models except the LR, the optimizers SGD and Adam are examined. For the LR, the learning rate scheduler \eqref{eq:lr-scheduler} along with a constant learning rate is investigated.

The best combination of hyperparameters is chosen as the set that minimized the mean absolute percentage error (MAPE) across the wells for each model type. The resulting hyperparameters for Case 1 and Case 2 can be seen in Tables \ref{app:hyperpar-search-all-data} and \ref{app:hyperpar-search-septest}, respectively. 

\subsection{Results of Case 1}\label{sec:results-case-1}
\begin{figure}[h!]
\centering
\includegraphics[width=1.0\linewidth]{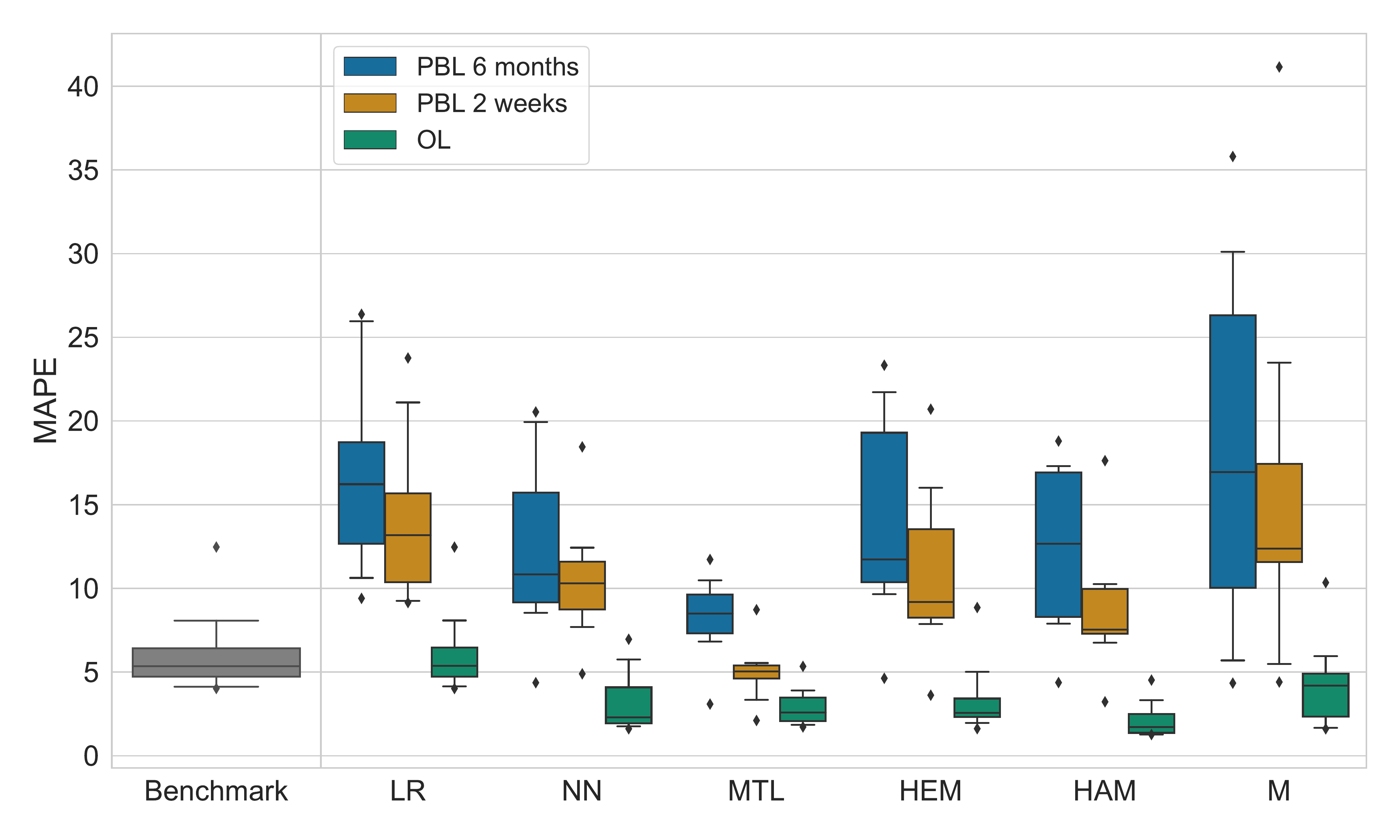}
\caption{The distribution of average error for each well, grouped for the models and learning methods. The models are trained with all available measurements. Compared to the performance of the benchmark model. The boxes show the $P_{25}$, $P_{50}$ (median), and $P_{75}$ percentiles. The whiskers show the $P_{10}$ and $P_{90}$ percentiles.}
\label{fig:mape-box-plot}
\end{figure}

In this case, both MPFM and well-test measurements are utilized in training. The box plot in Figure \ref{fig:mape-box-plot} shows the distribution of performances for the wells in terms of the MAPE grouped on the model type and learning method. The reported MAPE for one well is calculated using the predictions on all observations in the initial test set. The models are compared to the benchmark model. Table \ref{tab:MAPE-avg} gives an overview of the average MAPE across the wells for each model, and the last column presents the average MAPE of the learning methods across all wells and models. For the interested reader, Table \ref{app:MAPE-details} gives a detailed overview of the MAPEs for each well and model. There are several interesting observations. 

\begin{table}[h!]
\footnotesize
\caption{Average mean absolute percentage error across the wells for the models and learning methods trained on both MPFM and well-test measurements. The last column is the average MAPE across all wells and models.}
\begin{tabularx}{\textwidth}{lr*7{>{\raggedleft\arraybackslash}X}}
\toprule
Learning method & LR & NN & MTL & HEM & HAM & M & All \\
\midrule
PBL 6 months & 16.8 & 12.4 & 8.3 & 14.2 & 12.4 & 18.1 & 13.7\\
PBL 2 weeks & 14.2 & 10.5 & 5.0 & 10.9 & 8.7 & 15.7 & 10.8 \\
OL           & 6.2 & 3.2 & 2.9 & 3.4 & 2.1 & 4.2 & 3.7 \\
\bottomrule \noalign{\smallskip}
\end{tabularx}
\label{tab:MAPE-avg}
\end{table}

Firstly, as expected, the results clearly show that the model error decreases with an increased update frequency. On average, all models achieve a lower prediction error with PBL every second week compared to PBL every six months. With the OL, the average error decreases further with all models achieving an average error of less than 7\%. The overall best average performance across wells is achieved with OL on the HAM. The low MAPEs indicate that with access to frequently arriving measurements such as MPFM measurements, and allowed to learn continuously from them, the learning problem is relatively simple and a complex model is not necessary to achieve high VFM accuracy. This is supported by the good performance of the Benchmark which outperforms nearly all models trained with PBL. On the other hand, a disadvantage with the Benchmark is that it cannot be used for sensitivity analyses or in production optimization. 

\begin{figure}[h!]
\centering
\includegraphics[width=1.0\linewidth]{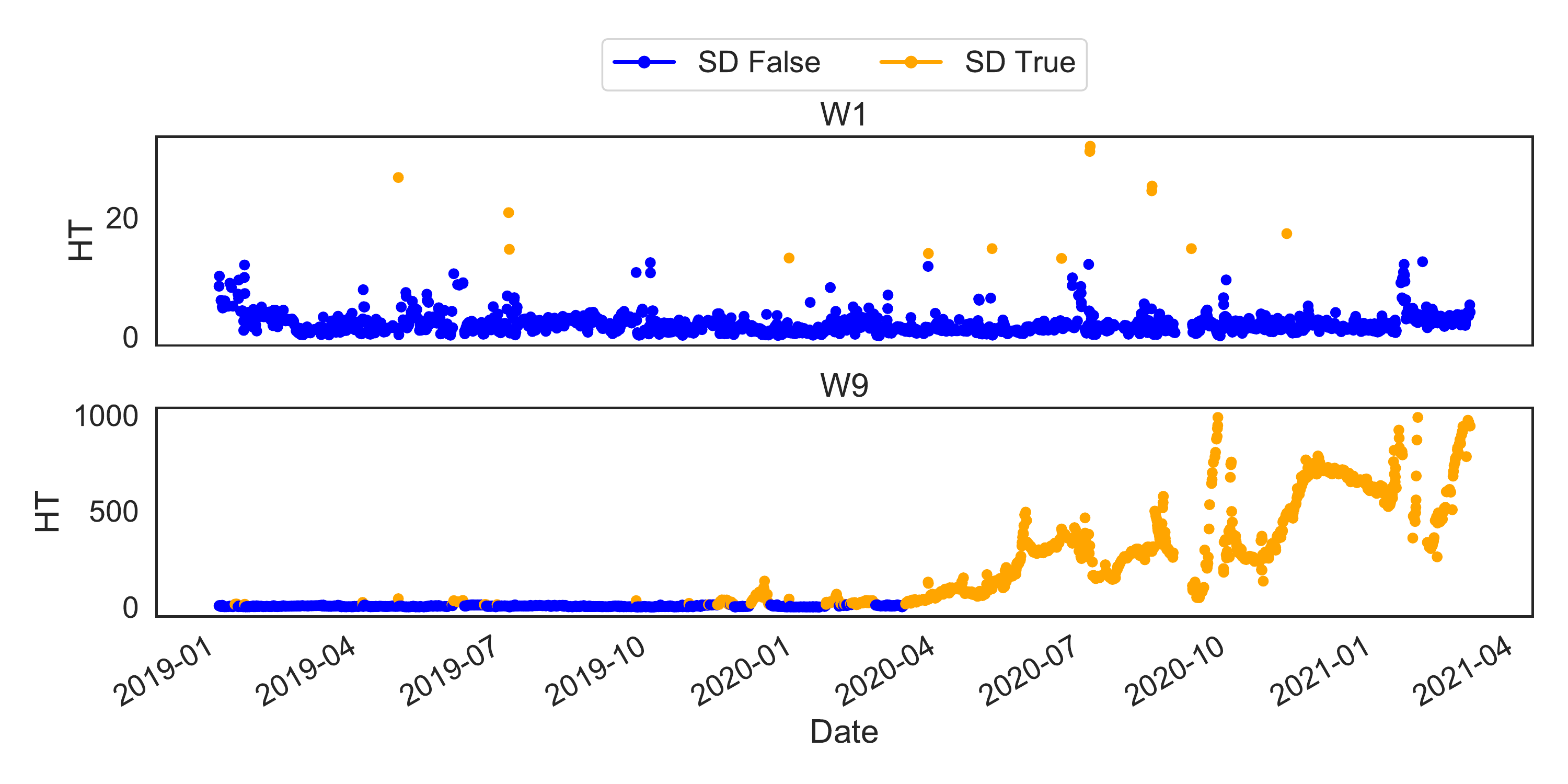}
\caption{The Hotelling's T-squared statistic for W1 and W9 comparing each observation from 01.01.2019 and forward with time to the training data containing the historical data before 01.01.2019. As seen, for W1 most observations are not detected as shifts. Whereas for W9, all observations towards to end are marked as a shift.}
\label{fig:HT-all}
\end{figure}

Secondly, from Table \ref{app:MAPE-details} it is observed that there are large differences in the error reduction for each well when the update frequency is increased. For instance, for W9 and most models, the error is greatly reduced going from the PBL 6 months to the OL. On the other hand, for W1 the reduction is not as prominent. This is likely related to whether or not the data generating distribution shifts with time. In Figure \ref{fig:HT-all}, the Hotelling's T-squared statistic is plotted for W1 and W9 using Algorithm \ref{alg:update-freq} on the initial training and test data. 
Figure \ref{fig:HT-all} indicates that it is unlikely that W1 experiences dataset shifts. On the other hand, for W9 it can be observed that the data likely shifts with time. Therefore, the results in Table \ref{app:MAPE-details} indicate that OL is better at tracking the local optimum of the learning problem when it changes with time. 

\begin{figure}[h!]
\centering
\includegraphics[width=1.0\linewidth]{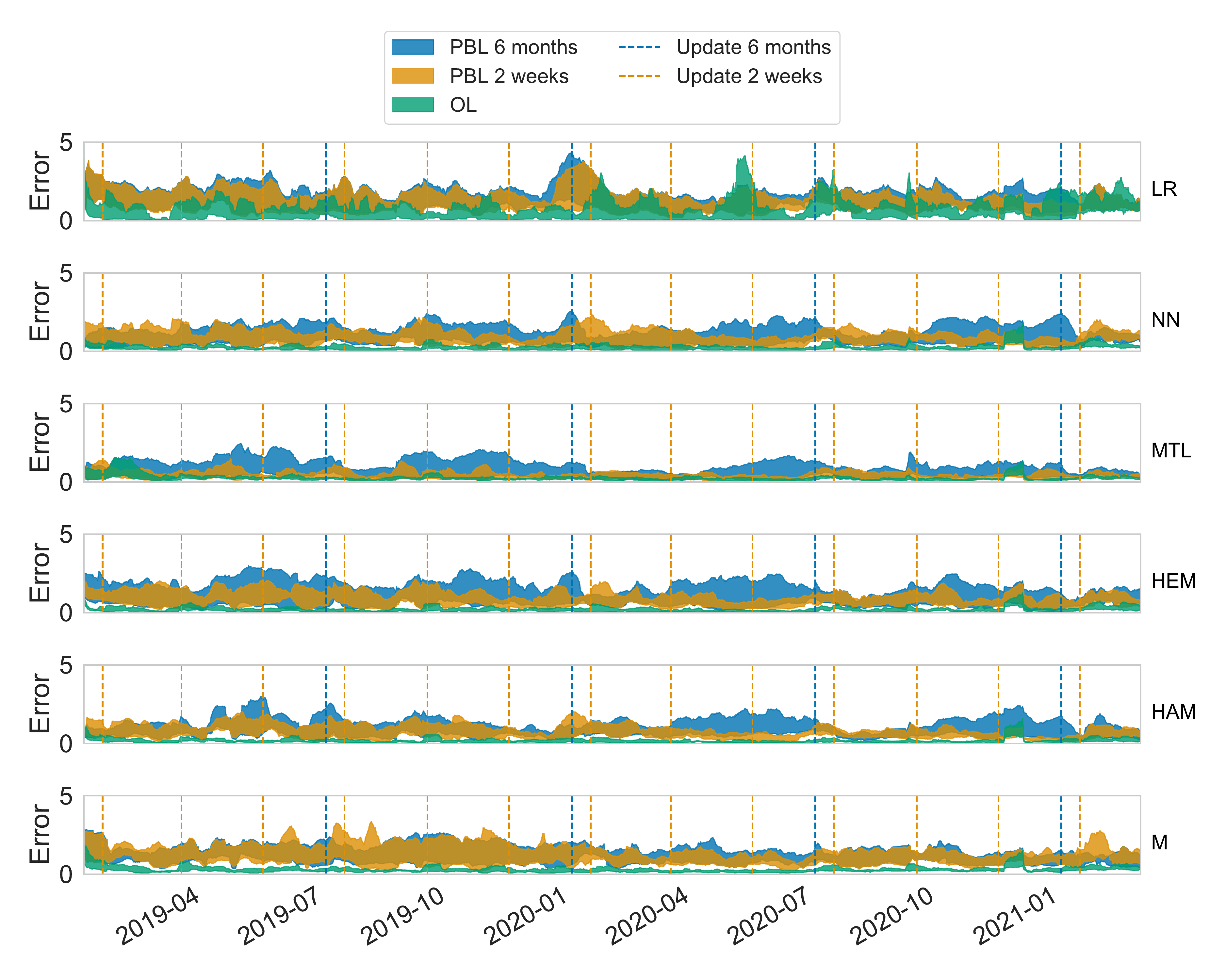}
\caption{The rolling absolute mean error across the wells against time for the models and learning methods. The window size used to calculate the error is 14 days. The shaded region illustrates the 25 and 75 percentiles of the errors across the wells. The vertical lines illustrate where the models are updated for the PBL 6 months and PBL 2 weeks.}
\label{fig:rolling_error}
\end{figure}

Another figure that illustrates the benefit of updating the model more often is Figure \ref{fig:rolling_error}, where the prediction error is visualized against time. 
The error is calculated as a rolling absolute mean error with a window size of 14 days. The shaded regions visualize the 25 and 75 percentiles of the errors across the wells. Notice that the PBL seems to yield a cyclic high and low accuracy. The average error increases with time up until model updating where the average error is reduced, naturally after some delay due to the rolling window. This is best observed for the PBL 6 months, but also to some extent for the PBL 2 weeks.   

\subsection{Results of Case 2}\label{sec:results-case-2}
\begin{figure}[h!]
\centering
\includegraphics[width=1.0\linewidth]{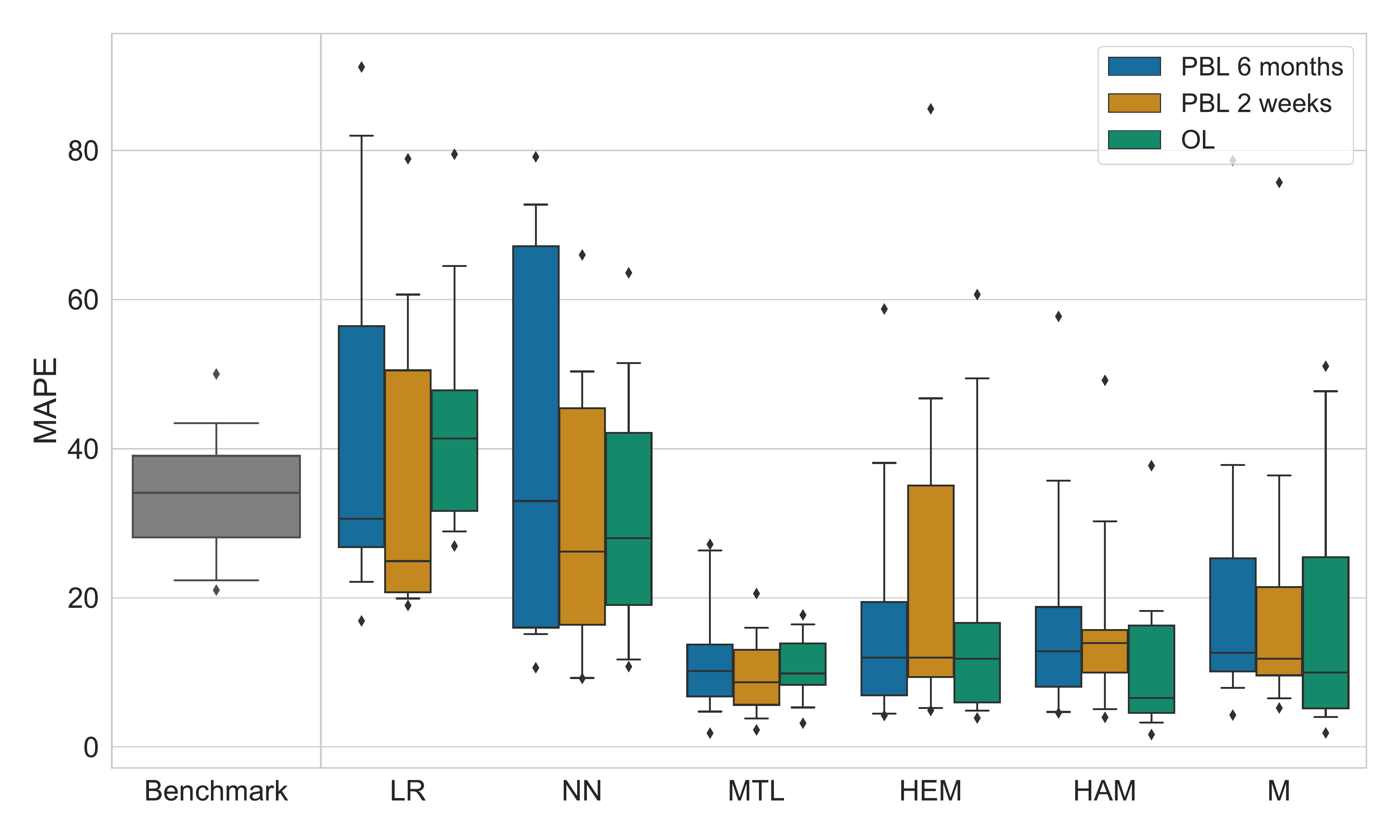}
\caption{The distribution of mean absolute percentage error (MAPE) for each well, grouped for the models and learning methods. Here the models are trained using well-test measurements only. Compared to the performance of the benchmark model. The boxes show the $P_{25}$, $P_{50}$ (median), and $P_{75}$ percentiles. The whiskers show the $P_{10}$ and $P_{90}$ percentiles.}
\label{fig:mape-box-plot-septest}
\end{figure}

\begin{figure}[h!]
\centering
\includegraphics[width=1.0\linewidth]{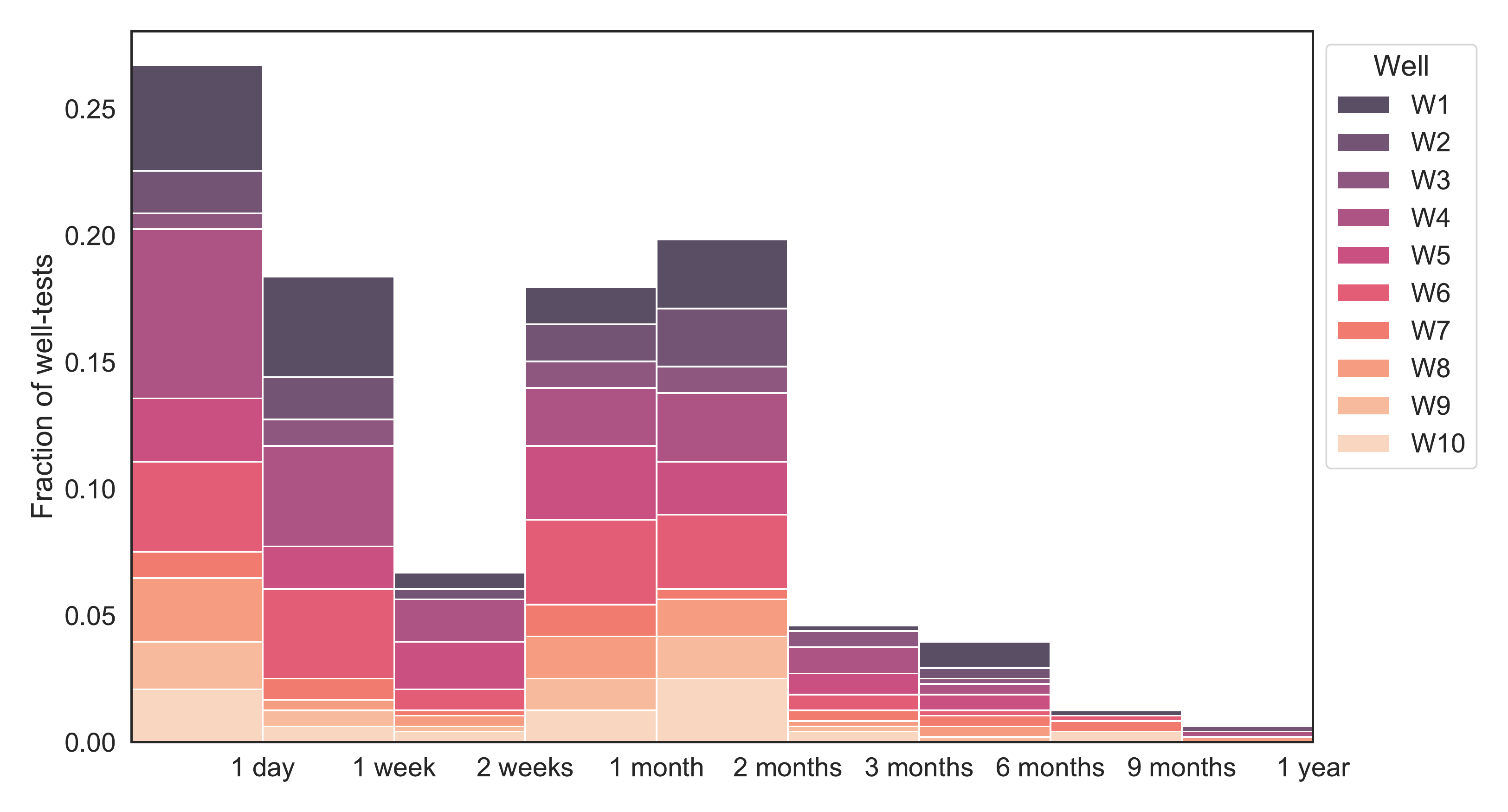}
\caption{The elapsed time between new well-tests for each well present in the dataset. Pay in mind that the bins in the histogram have different sizes, from one day to three months of elapsed time. Observe that many of the wells have measurements occurring more than two weeks apart.}
\label{fig:time-between-welltest}
\end{figure}

In this case, the models are trained on well-test measurements only, see the observations colored orange in Figure \ref{fig:dataset_visualization}. Figure \ref{fig:mape-box-plot-septest} illustrates the distribution of MAPEs for the wells. Table \ref{tab:MAPE-avg-septest} gives an overview of the average MAPE across the wells. Table \ref{app:MAPE-details-septest} reports the MAPE for each well, model, and method. 
\begin{table}[h!]
\footnotesize
\caption{The average mean absolute percentage error across the wells for the models and learning methods trained on well-test measurements. The last column is the average MAPE across all wells and models.}
\begin{tabularx}{\textwidth}{lr*6{>{\raggedleft\arraybackslash}X}}
\toprule
Learning method & LR & NN & MTL & HEM & HAM & M & All \\
\midrule
PBL 6 months & 43.2 & 40.7 & 12.1& 17.9 & 18.3 & 21.8 & 25.7 \\
PBL 2 weeks & 37.1 & 31.1 & 9.6 & 24.5 & 16.7  & 20.5 & 23.3 \\
OL           & 44.3 & 31.4 & 10.5 & 18.7 & 11.6 & 17.7 & 22.4 \\
\bottomrule \noalign{\smallskip}
\end{tabularx}
\label{tab:MAPE-avg-septest}
\end{table}
First of all, notice the significantly different results obtained for this case compared to Case 1. In Case 1, a trend of decreased error for increased update frequency is observed. Here, the difference in performance is negligible for many models and for other models the error increases going from PBL to OL. The observed results are likely related to the elapsed time between each new well-test, illustrated in Figure \ref{fig:time-between-welltest} by a stacked histogram. Notice that many of the wells have several tests that are more than a month apart. Furthermore, eight of ten wells have the majority of tests occurring with a frequency lower than 14 days, see Table \ref{app:time-between-tests}. In such situations, the frequency of model updating is equal for PBL 2 weeks and OL, and the only difference between the two is how the updating is executed. The low frequency of well-tests is also likely the cause of the decreased Benchmark performance compared to Case 1. With a lower frequency, the process conditions can have changed significantly in-between well-tests and two chronological flow rate measurements are likely uncorrelated. The intermittent time between well-tests also makes it challenging to obtain good hyperparameters. If well-tests occur frequently, the model will likely require small parameter updates, and opposite for infrequently arriving well-tests. Non-optimal hyperparameters can explain the overall poorer average performance for all models and methods than for Case 1.

The large MAPEs in Table \ref{tab:MAPE-avg-septest} show that in the presence of infrequent and intermittent measurements, the learning problem is not trivial and a more complex model than, for instance, the Benchmark, is required to obtain an adequate performance. Nevertheless, the comparable performance of the LR and NN indicate that choosing a data-driven model with higher complexity is not the solution to increased performance in this case. Likely, the amount of data available is too small for high-capacity data-driven models to exploit their capacity. An observation that supports this is that the NN obtained a significant improved performance in Case 1 where the amount of data is higher. On the other hand, having physical considerations in the model structure does seem to be be advantageous for VFM when the amount of data is small. From Figure \ref{fig:mape-box-plot-septest}, the MTL, HEM, HAM, and M, all achieve a median MAPE below 20\% whereas the NN and LR are well above 20\%. Another interesting observation is that learning from several wells as for the MTL seems to yield a more robust approach as the spread in performances is low.  

\section{Concluding remarks}\label{sec:conclusions}
The results in this research show that a high update frequency of steady-state VFM models is key to sustaining a high performance in nonstationary conditions over time. In particular, if the frequency of measurement arrival is high. Therefore, for petroleum assets with access to multiphase flow meters, steady-state VFMs can yield an excellent performance. Of the two passive learning methods analyzed, online learning achieves the best average performance with an error of 3.7\% across all wells and model types. This is an error reduction of 73\% compared to the average error of periodic batch learning with an update frequency of 6 months. On the other hand, if the arrival of new measurements is intermittent and with low frequency, which is a common issue on assets with well-testing only, the benefit of frequent model updating is small and less evident. This is likely due to the challenging task of finding good hyperparameters. The average error increased significantly with online learning achieving an average error of 22.4\% across all wells and model types. However, an interesting observation is that VFM model types with physical considerations seem to offer the best performance in the presence of little data. 

Hence, the results show that online learning seems a promising method to obtain high accuracy steady-state VFM models, in particular, with frequently arriving measurements. However, the method will also require fast measurement processing capabilities, model performance monitoring applications, and automatic handling of the learning process. Therefore, online learning can be challenging to integrate into existing systems. Likely, an appropriate learning method must be chosen as a trade-off between accuracy and available resources. With limited resources, periodic batch learning with frequent model updating, for instance, every second week, can be better suited in real-time applications. 

Although the learning methods in this research are investigated for 10 typical subsea wells on the Norwegian continental shelf, these are certainly not representative for all production wells as the multiphase flow characteristics can be very different. Therefore, it is hard to generalize the results and it would benefit the conclusion if more wells from different assets are included. Nevertheless, the overall conclusion of this research is that passive learning with frequent model updating can significantly improve the accuracy of steady-state VFMs in nonstationary environments. The investigation can be of interests to experts developing soft-sensors, like VFMs.  

\section*{Acknowledgments}
The authors would like to thank Lundin Energy Norway for allowing them to work with production data from a real petroleum field. They would further like to thank Solution Seeker AS for the contribution with data collection and pre-processing.

\section*{Funding}
This research is supported by Lundin Energy Norway and is a part of BRU21 - NTNU Research and Innovation Program on Digital and Automation Solutions for the Oil and Gas Industry (www.ntnu.edu/bru21). Lundin Energy Norway has not taken part in data collection and analysis, nor in writing of the report. However, the article is approved by Lundin before submission for publication. 

\clearpage
\appendix
\section{Estimation of the update frequency in periodic batch learning}\label{app:update-frequency-pbl}
Consider the null hypothesis $\mathcal{H}_0$ to state that there is no virtual drift present in the data such that input distribution $p(\bm{x})$ does not shift with time. This $\mathcal{H}_0$ is also called the stationary hypothesis. The $\mathcal{H}_1$ hypothesis is the alternative hypothesis that there is a shift in the data. Mathematically:
\begin{equation}
\begin{aligned}
    \mathcal{H}_0&: \quad p_{t}(\bm{x}) = p_{t + \tau}(\bm{x}) \quad \text{for all } \tau > 0 \\
    \mathcal{H}_1&: \quad p_{t}(\bm{x}) \neq p_{t + \tau}(\bm{x}) \quad \text{for any } \tau > 0.
\end{aligned} 
\end{equation}

Consider two disjoint datasets $\dataset_{1}$ and $\dataset_{2}$ with size $N_{1}$ and $N_{2}$ and inputs observations $\bm{X}_1 \in \reals^{d\times N_1}$ and $\bm{X}_2 \in \reals^{d\times N_2}$, respectively. The Hotelling's T-squared statistic calculates the probability of equal means of the two multivariate input distributions at a significance level $\alpha$. The statistic is calculated as
\begin{equation}\label{eq:hotellings}
    HT^2 = (\bm{\mu}_1 - \bm{\mu}_2)\left( \frac{\bm{\Sigma}_1}{N_1} + \frac{\bm{\Sigma}_2}{N_2} \right)^{-1} (\bm{\mu}_1 - \bm{\mu}_2)^\top, 
\end{equation}
where $\bm{\mu} \in \reals^{d}$ is the sample mean vector and the $\bm{\Sigma}\in\reals^{d\times d}$ is the sample covariance matrix of the input. The Hotelling's T-squared statistic follows the F-distribution $F(d, N_1 + N_2 - d - 1)$ \citep{Hardle2012}. To estimate an appropriate update frequency in PBL, the two-sample Hotelling's T-squared test can be used on available training data using Algorithm \ref{alg:update-freq}

\begin{algorithm}[ht]
\caption{Estimation of the update frequency $\tau$ in periodic batch learning}
\label{alg:update-freq}
\small
\begin{algorithmic}[1]
\Require data $\dataset_{1:T} = \{(\bm{x}_t, y_t)\}_{t=1}^{T}$, significance level $\alpha$
\State Set $\dataset_{1} = \dataset_{1:T_1}$ where $1<T_1<T$. 
\For{$k=1,...,T-T_1$}
    \State $\dataset_{2,k} = \dataset_{T_1+k}$
    \State Calculate $HT^2$ using \eqref{eq:hotellings} with $\dataset_1$ and $\dataset_{2,k}$
    \State Calculate F-statistic $F_k$ for $HT^2$
    \State Calculate the critical value $F_{\text{crit}}$ at significance level $\alpha$
        \If{$F_k < F_{\text{crit}}$}
            \State Reject $\mathcal{H}_0$, shift detected
            \State \Return $\tau = T_1 + k$
        \EndIf
\EndFor
\end{algorithmic}
\end{algorithm}
The algorithm is subject to false shift detections, or type II error, for instance, if the observation is faulty or noisy. A workaround is to test additional observations following with time. If shifts are detected on several subsequent observations, virtual drift has likely occurred. If the following observations are not detected as shifts, likely, the detection is falsely reported. 

\section{Hyperparameter search}
\setcounter{table}{0}
\renewcommand{\thetable}{B\arabic{table}}
\begin{table}[h!]
\tiny
\caption{The training algorithm settings as a result of the hyperparameter search for training on both MPFM and well-test measurements. For the batch learning approaches, only the value of the learning rate $\gamma$ is experimented with. The number of iterations $E$ for PBL is found with early stopping (E.S.). The (s.) and (c.) for the LR refers to using the learning rate scheduler in \eqref{eq:lr-scheduler} and constant learning rate, respectively.}
\begin{tabularx}{\textwidth}{lr*9{>{\raggedleft\arraybackslash}X}}
\toprule
Model & \multicolumn{3}{c}{PBL 6 months} & \multicolumn{3}{c}{PBL 2 weeks}  & \multicolumn{3}{c}{OL} \\
& $\gamma$ & $E$ & $O$ & $\gamma$ & $E$ & $O$ & $\gamma$ & $E$ & $O$ \\
\midrule 
LR & (s.) $10^{-2}$ & E.S. & SGD & (s.) $10^{-1}$  & E.S. & SGD & (c.) $0.5$  & 20 & SGD \\ 
NN & $10^{-4}$ & E.S. & Adam & $10^{-3}$ & E.S. & Adam & $10^{-5}$ & 20 & Adam \\ 
MTL & $10^{-4}$ & E.S. & Adam & $10^{-3}$ & E.S. & Adam & $10^{-6}$ & 20 & Adam \\
HEM & $10^{-3}$ & E.S. & Adam & $10^{-3}$ & E.S. & Adam & $10^{-2}$ & 20 & SGD \\
HAM & $10^{-3}$ & E.S. & Adam & $10^{-3}$ & E.S. & Adam & $10^{-5}$ & 20 & SGD\\ 
M & $10^{-3}$ & E.S. & Adam & $10^{-3}$ & E.S. & Adam & $10^{-2}$ & 10 & Adam\\
\bottomrule \noalign{\smallskip}
\end{tabularx}
\label{app:hyperpar-search-all-data}
\end{table}

\begin{table}[h!]
\tiny
\caption{The training algorithm settings as a result of the hyperparameter search for training on only well-test measurements. For the batch learning approaches, only the learning rate $\gamma$ is experimented with. The number of iterations $E$ for PBL is found with early stopping (E.S.). The (s.) and (c.) for the LR refers to using the learning rate scheduler in \eqref{eq:lr-scheduler} or constant learning rate, respectively.}
\begin{tabularx}{\textwidth}{lr*9{>{\raggedleft\arraybackslash}X}}
\toprule
Model & \multicolumn{3}{c}{PBL 6 months} & \multicolumn{3}{c}{PBL 2 weeks}  & \multicolumn{3}{c}{OL} \\
& $\gamma$ & $E$ & $O$ & $\gamma$ & $E$ & $O$ & $\gamma$ & $E$ & $O$ \\
\midrule 
LR & (s.) $10^{-4}$  & E.S. & SGD & (s.) $0.5$  & E.S. & SGD & (s.) $10^{-3}$ & 1 & SGD \\ 
NN & $10^{-4}$ & E.S. & Adam & $10^{-3}$ & E.S. & Adam & $10^{-4}$\ph & 20 & SGD \\ 
MTL & $10^{-5}$ & E.S. & Adam & $10^{-5}$ & E.S. & Adam & $10^{-5}$\ph & 20 & Adam \\
HEM & $10^{-3}$ & E.S. & Adam & $10^{-4}$ & E.S. & Adam & $10^{-10}$ & 20 & SGD \\
HAM & $10^{-5}$ & E.S. & Adam & $10^{-5}$ & E.S. & Adam & $10^{-5}$\ph & 20 & Adam\\ 
M & $10^{-3}$ & E.S. & Adam & $10^{-3}$ & E.S. & Adam & $10^{-2}$\ph & 10 & Adam\\
\bottomrule \noalign{\smallskip}
\end{tabularx}
\label{app:hyperpar-search-septest}
\end{table}

\clearpage
\section{Additional results from the numerical study}
\setcounter{table}{0}
\renewcommand{\thetable}{C\arabic{table}}
\begin{table}[h!]
\tiny
\caption{Mean absolute percentage errors for all wells and models. The triple of numbers reported is the error for periodic batch learning every 6 months, periodic batch learning every 2 weeks, and online learning, respectively. }
\begin{tabularx}{\textwidth}{lr*6{>{\raggedleft\arraybackslash}X}}
\toprule
Well & LR & NN & MTL & HEM & HAM & M \\
\midrule 
1 & 9.4, \ph9.3, 12.5 & 4.4 \ph4.9, 1.6 & 3.1, 2.1, 1.7 & 4.6, \ph3.6, 2.0 & 4.4, \ph3.2, 1.3 & 4.3, \ph4.4, \ph1.7 \\ 
2 & 26.0, 21.1, \ph5.4 & 16.4, 18.5, 5.8 & 10.5, 8.7, 5.3 & 19.4, 20.7, 8.9 & 18.8, 17.6, 4.5 & 35.8, 41.2, 10.3  \\ 
3 & 12.6, \ph9.8, \ph4.0 & 10.0, 11.7, 7.0 & 9.8, 4.9, 3.9 & 10.9, \ph8.8, 5.0 & 14.7, 10.2, 2.7 & 30.1, 11.7, \ph5.9 \\  
4 & 12.8, 12.0, \ph8.1 & 10.4, \ph9.8, 1.8 & 7.2, 5.0, 1.8 & 10.5, \ph9.2, 2.4 & 8.1, \ph7.3, 1.3 & 5.7, \ph5.5, \ph2.3 \\  
5 & 18.6, 16.0, \ph6.7  & 8.9, \ph8.5, 3.2 & 7.5, 3.3, 2.1 & 12.6, \ph9.2, 1.6 & 10.6, \ph7.6, 1.4 & 13.5, 11.5, \ph1.6 \\  
6 & 17.8, 14.8, \ph5.3 & 13.7, 11.1, 2.4 & 8.6, 5.4, 3.6 & 19.0, 16.0, 2.3 & 17.1, 10.3, 1.7 & 27.4, 23.5, \ph4.1\\  
7 & 26.4, 23.8, \ph5.7 & 20.5, 12.4, 4.4 & 11.7, 5.1, 3.1 & 23.3, 11.9, 2.9 & 17.3, \ph9.1, 3.3 & 20.4, 16.5, \ph4.3 \\  
8 & 14.7, 12.6, \ph4.1 & 11.2, 10.8, 1.9 & 6.8, 5.5, 2.5 & 9.6, \ph7.9, 2.7 & 7.9, \ph6.7, 1.7 & 11.2, 12.8, \ph5.0 \\
9 & 10.6, \ph9.2, \ph5.3 & 8.5, \ph7.7, 2.2 & 8.3, 5.3, 2.1 & 10.3, \ph8.1, 1.7 & 8.9, \ph7.4, 1.9 & 9.7, 12.0, \ph2.6 \\  
10 & 18.8, 13.7, \ph4.5 & 19.9, \ph9.4, 2.2 & 9.2, 4.5, 2.7 & 21.7, 14.1, 3.6 & 16.4, \ph7.3, 1.9 & 23.0, 17.8, \ph4.5\\  
\bottomrule \noalign{\smallskip}
\end{tabularx}
\label{app:MAPE-details}
\end{table}

\begin{table}[h!]
\tiny
\caption{Mean absolute percentage errors for all wells and models trained on only test separator measurements. The triple of numbers reported is the error for periodic batch learning every 6 months, periodic batch learning every 2 weeks, and online learning, respectively. }
\begin{tabularx}{\textwidth}{lr*6{>{\raggedleft\arraybackslash}X}}
\toprule
Well & LR & NN & MTL & HEM & HAM & M \\
\midrule 
1 & 26.5, 19.9, 28.9 & 10.6, \ph9.3, 10.8 & 1.9, \ph2.3, \ph3.2 & \ph4.2, \ph5.2, \ph7.9 & \ph4.6, \ph4.0, \ph1.7 & \ph4.3, \ph6.5, \ph1.9\\ 
2 & 91.2, 78.8, 64.5 & 62.0, 66.0, 30.3 & 27.2, 16.0, 17.7 & 58.7, 85.6, 49.4 & 57.7, 49.2, 37.7 & 78.7, 75.7, 47.7\\ 
3 & 22.2, 20.0, 46.9 & 79.1, 25.4, 63.6 & 11.1, 10.7, 14.7 & 21.6, 46.8, 17.6 & 14.1, 14.7, 16.5 & 28.9, 23.8, 51.1\\  
4 & 32.9, 24.4, 34.8 & 16.6, 14.7, 18.5 & 11.0, 10.1, \ph 8.4 & 12.8, 10.6, \ph5.3 & 10.8, \ph9.9, \ph7.0 & \ph9.9, 10.5, \ph5.8\\  
5 & 28.3, 25.4, 26.9 & 15.1, \ph9.2, 11.7 & 4.7, \ph3.8, \ph 5.3 & \ph4.5, \ph4.9, \ph3.9 & \ph7.2, 10.2, \ph4.3 & \ph7.9, \ph5.2, \ph4.9\\  
6 & 82.0, 60.7, 48.1& 72.7, 47.0, 43.7 & 14.6, 13.8, 11.4 & \ph8.6, 11.8, 13.3 & 11.6, 13.9, \ph5.4 & 14.5, 13.1, \ph8.8\\  
7 & 43.5, 50.8, 79.5 & 68.9, 50.4, 51.5 & 9.3, \ph6.4, 11.2 & 11.7, 18.2, 13.7 & 19.1, 16.0, 18.2 & 10.8, 14.2, 11.6\\  
8 & 16.9, 19.0, 30.6 & 15.8, 21.3, 20.8 & 6.3, \ph5.4, \ph 8.3 & 6.4, 12.2, 10.4 & \ph4.7, \ph5.1, \ph6.2 & 12.8, \ph9.5, 11.1\\
9 &27.7, 22.9, 36.7  & 27.1, 27.0, 25.7 & 8.1, \ph7.2, \ph 8.5 & 12.3, \ph9.0, \ph4.9 & 17.6, 14.0, \ph3.3& 12.5, \ph9.9, \ph4.0\\  
10 &60.7, 49.8, 46.1 & 38.9, 40.5, 37.8 & 26.3, 20.6, 16.4 & 38.1, 40.7, 60.7 & 35.7, 30.3, 15.6 & 37.8, 36.4, 30.1\\  
\bottomrule \noalign{\smallskip}
\end{tabularx}
\label{app:MAPE-details-septest}
\end{table}

\begin{table}[h!]
\tiny
\caption{The percentage of well-tests for a well where the number of days between two chronological tests resides in the given bins. d=day, w=week, m=month, y=year.}
\begin{tabularx}{\textwidth}{lr*7{>{\raggedleft\arraybackslash}X}}
\toprule
Well & < 2w & 2w-1m & 1m-2m & 2m-3m & 3m-6m & 6m-9m & 9m-1y\\
\midrule 
1 & 60.9\% & 10.1\% & 18.8\%\ & 1.5\% & 7.2\% & 1.5\% & 0\%\\ 
2 & 46.2\% & 17.9\% & 28.2\% & 0\% & 5.1\% & 0\% & 2.6\% \\
3 & 36.4\% & 22.7\% & 22.7\% & 13.6\% & 4.5\% & 0\% & 0\%  \\
4 & 64.8\% & 12.1\% & 14.3\% & 5.5\% & 2.2\% & 0\% & 1.1\%  \\
5 & 48.3\% & 23.3\% & 16.7\% & 6.7\% & 5.0\% & 0\% & 0\%  \\
6 & 52.1\% & 21.9\% & 19.2\% & 4.1\% & 1.4\% & 1.4\% & 0\%  \\
7 & 41.7\% & 25.0\% & 8.3\% & 8.3\% & 8.3\% & 8.3\% & 0\%  \\
8 & 45.7\% & 22.9\% & 20.0\% & 2.9\% &5.7\% & 0\% & 2.9\%  \\
9 & 44.8\% & 20.7\% & 27.6\% & 3.4\% & 3.4\% & 0\% & 0\%  \\
10 & 40.5\% & 16.2\% & 32.4\% & 5.4\% & 0\% & 5.4\% & 0\%\\
\bottomrule \noalign{\smallskip}
\end{tabularx}
\label{app:time-between-tests}
\end{table}

\clearpage 
\bibliography{bibliography.bib}

\end{document}